\documentclass[lettersize,journal]{IEEEtran}

\usepackage{soul}
\usepackage{amssymb, amsmath, amsthm}
\usepackage{mathtools}
\usepackage{bm}
\usepackage{algorithm}
\usepackage{algpseudocode}
\usepackage{graphicx}
\usepackage{subfigure}
\usepackage{multirow}
\usepackage{longtable}
\usepackage{graphicx}
\usepackage{siunitx}
\usepackage{booktabs}
\usepackage{mathtools}
\usepackage{caption}
\usepackage{graphicx,xcolor}
\usepackage[framemethod=tikz]{mdframed}
\usepackage{comment}
\usepackage{graphicx}
\usepackage{mathabx}
\usepackage{hyperref}
\usepackage{enumitem}

\newcommand\scalemath[2]{\scalebox{#1}{\mbox{\ensuremath{\displaystyle #2}}}}
\newcommand{\ubar}[1]{\text{\b{$#1$}}}

\newcommand\norm[1]{\lVert#1\rVert}
\algnewcommand{\IIf}[1]{\State\algorithmicif\ #1\ \algorithmicthen}
\algnewcommand{\EndIIf}{\unskip\ \algorithmicend\ \algorithmicif}

\DeclareMathOperator{\Int}{int}
\DeclareMathOperator{\Sat}{sat}

\DeclareMathOperator{\diag}{diag}

\newcommand{\REV}[1]{#1}
\newcommand{\REVV}[1]{#1}

\DeclareMathOperator{\train}{tr}
\DeclareMathOperator{\val}{val}
\DeclareMathOperator{\test}{te}

\newtheorem{definition}{Definition}
\newtheorem{lemma}{Lemma}
\newtheorem{theorem}{Theorem}
\newtheorem{remark}{Remark}
\newtheorem{property}{Property}
\newtheorem{assumption}{Assumption}
\newtheorem{proposition}{Proposition}

\begin{document}
\title{\REV{Learning Control Affine Neural NARX Models for Internal Model Control design}}

\author{Jing Xie$^\dagger$, Fabio Bonassi, Riccardo Scattolini
\thanks{$^\dagger$ Corresponding author.}
\thanks{Jing Xie and Riccardo Scattolini are with Dipartimento di Elettronica Informazione e Bioingegneria, Politecnico di Milano, Italy. Email: \texttt{jing.xie{@}polimi.it}, \texttt{riccardo.scattolini{@}polimi.it}} 
\thanks{Fabio Bonassi is with the Department of Information Technology, Uppsala University, Sweden. Email: \texttt{fabio.bonassi@it.uu.se}}
\thanks{This project has received funding from the European Union’s Horizon 2020 research and innovation programme under the Marie Skłodowska-Curie grant agreement No. 953348.}}
\maketitle

\begin{abstract}
This paper explores the use of Control Affine Neural Nonlinear AutoRegressive eXogenous (CA-NNARX) models for nonlinear system identification and model-based control design.
The idea behind this architecture is to match the known control-affine structure of the system to achieve improved performance.
Coherently with recent literature of neural networks for data-driven control, we first analyze the stability properties of CA-NNARX models, devising sufficient conditions for their incremental Input-to-State Stability ($\delta$ISS) that can be enforced at the model training stage.
The model's stability property is then leveraged to design a stable Internal Model Control (IMC) architecture.
The proposed control scheme is tested on a real Quadruple Tank benchmark system to address the output reference tracking problem.
The results achieved show that (\textit{i}) the modeling accuracy of CA-NNARX is superior to the one of a standard NNARX model for given weight size and training epochs,  (\textit{ii}) the proposed IMC law provides performance comparable to the ones of a standard Model Predictive Controller (MPC) at a significantly lower computational burden\REV{, and (\textit{iii}) the $\delta$ISS of the model is beneficial to the closed-loop performance.}
\end{abstract}
\def\abstractname{Note to Practitioners}
\begin{abstract}
Many engineering systems, such as robotic manipulators and chemical
reactors, are described by Control Affine (CA) models, characterized by onlinear dynamics where the control variable enters in a linear way. If only this structural information is available without any additional knowledge, for instance on the order of the system or on the value of its parameters, a black-box identification approach can be followed to estimate the model from data. For these reasons, in this paper we propose a modeling and control design method suited for this class of systems. Specifically, we assume that the system is described by a CA-Neural Nonlinear AutoRegressive eXogenous (CA-NNARX) model. Then, the estimated model is used to design a stable Internal Model Control (IMC) scheme for the solution of output reference tracking problems. The stability, performance, and robustness properties of the proposed approach are studied and tested in the control of a laboratory system.
In addition, a simulation analysis shows how IMC represents a valid alternative to the popular Model Predictive Control (MPC) approach, in particular for embedded systems, where the computation power required by MPC can be too high.
\end{abstract}


\section{Introduction}
\REV{
Learning-based and data-driven methods are playing an increasingly important role in the design of control and monitoring systems of industrial plants. This is motivated by ongoing technological developments, which allow for the availability of more and more local computing capabilities and information about the system in terms of measurements. 
In parallel, new and efficient machine learning methods for control design are being developed for a wide set of applications. Among them, the most recent direct control synthesis approaches for linear systems are based on the celebrated Willems lemma, see \cite{berberich2024overview} and references therein. In a nonlinear setting, it is worth recalling the methods based on Gaussian Processes, \cite{hewing2019cautious}, Koopman's operators, \cite{mauroy2020koopman}, and Neural Networks (NN), \cite{bonassi2022recurrent}. In particular, NN are used in a variety of approaches, due to their high flexibility and performance as functional approximators. As discussed in  \cite{bonassi2022recurrent}, NN are the basis of many direct and indirect data-driven control architectures, and might also be employed, for example, to approximate control laws computed off-line, or in the framework of adaptive control \cite{liu2021adaptive}, reinforcement learning \cite{perrusquia2021identification}, or fuzzy neural control methods \cite{bu2022low, bu2023fuzzy, bu2021simplified}.}
\REV{Using NN as nonlinear identification models of complex systems opens the way to the design of sophisticated model-based control synthesis methods, like Internal Model Control (IMC), see \cite{garcia1982internal, saxena2012advances, economou1986internal} or Model Predictive Control (MPC), see \cite{camacho2007model, mayne2000constrained, rawlings2017model}, which are nowadays among the most widely used advanced control methods in the process industry.}

\smallskip
\REV{Despite the impressive results achieved in many control problems, however, NN are still treated with some caution by the control community due to their difficult interpretability, and to the lack of solid theoretical guarantees concerning stability and generalization, i.e. the capability to produce meaningful and consistent predictions even for data not included in the training set. To partially overcome these limitations, one of the most promising fields of research concerns the development of conditions guaranteeing stability properties for the most popular families of Recurrent Neural Networks (RNN), such as Echo State Networks (ESN, \cite{jaeger2007echo}), Gated Recurrent Units (GRU, \cite{chung2014empirical}), Long Short-Term Memory networks (LSTM, \cite{hochreiter1997long}), \cite{schimperna2024robust}), and  Neural Nonlinear AutoRegressive eXogenous networks (NNARX, \cite{levin1996control}). 
For all these families, conditions on their parameters guaranteeing the Input-to-State Stability (ISS, \cite{jiang2001input}) and Incremental Input-to-State Stability ($\delta$ISS, \cite{bayer2013discrete}) have been established. 
Based on these conditions, training procedures have been developed for learning RNN with stability certifications, see \cite{bonassi2022recurrent} and the references therein.
Finally, stable RNN have been shown to allow for the design of  controllers achieving closed-loop stability guarantees \cite{chen2020machine}. Concerning this point, the reader is addressed to  \cite{bonassi2022offset, xie2022robust} for NNARX, \cite{bonassi2021nonlinear} for GRU, and \cite{terzi2019lstm, schimperna2023offset} for LSTM.}

\REV{
Besides stability, another problem that has been partially addressed is that of the interpretability of these RNN models, that is, the possibility to easily understand the meaning of parameters and structure of RNN.
To this end, some approaches have been recently developed, all of them sharing the basic idea to enforce known relations in the model structure definition, or among their variables, see \cite{willard2022integrating, ZHENG2023physics, WU202074pinn}. 
A common outcome of these method is that the estimated physics-informed models are more interpretable, physically consistent, and easy to train also in case of complex models, see for example \cite{willard2022integrating,boca2024physicsnn}.}

\subsection{Motivations and goals}
\REV{
In this paper, we consider the problem of controlling an unknown Control Affine (CA) system \cite{isidori1985nonlinear}, that is, a nonlinear system affine with respect to the input vector $u \in \mathbb{R}^{n_u}$.
Denoting by $x \in \mathbb{R}^{n_x}$ and $y \in \mathbb{R}^{n_y}$ the state and output vector, respectively, a CA system can be written as
\begin{equation} \label{eq:intro:ca}
\begin{dcases}
	x_{k + 1} = f(x_k) + g(x_k) u_k \\
	y_k = h(x_k)
\end{dcases}
\end{equation}
for some nonlinear Lipschitz-continuous function $f(\cdot)$, $g(\cdot)$, and $h(\cdot)$.
Many well-known engineering systems have this structure, such as the single-link flexible joint manipulators \cite{BANSAL2021375}, spacecraft systems \cite{chen2020space}, attitude control systems, \cite{wang2023time}, \cite{wang2022nonlinear}, and wheeled robotic systems \cite{wu2019robottracking}.
In the case where the dynamics of the system are not known, however, identifying a system in the form \eqref{eq:intro:ca} is generally difficult \cite{tang2022data}.}

\REV{
In the spirit of physics-informed machine learning, we here propose an architecture, henceforth referred to as CA-NNARX, that allows us to identify a model that enjoys the same control affine structure \eqref{eq:intro:ca} as the system to be identified.
This model is learned from the input-output data collected from the unknown plant by means of a tailored training procedure that allows to enforce its $\delta$ISS. This property, as discussed in \cite{bonassi2023reconciling}, allows for models that are stable and robust to perturbations.}

\REV{Once a CA-NNARX model of the plant is available, we tackle the problem of designing a model-based control scheme able to track output reference signals.
We propose to adopt an IMC procedure owing to its conceptually-simple design and its extremely low online computational cost.}

\subsection{Contributions}
\REV{The contributions to the literature are the following. 
First, we formalize CA-NNARX models, and derive a  sufficient condition on their learnable parameters under which the $\delta$ISS property can be guaranteed.
Such a condition is then enforced during the model's training stage to learn provenly $\delta$ISS CA-NNARX models.}

\REV{Second, based on the learned $\delta$ISS CA-NNARX model, we synthesize an IMC law. 
Notably, IMC calls for the model's inverse, which generally is not available in nonlinear settings and thus needs to be numerically approximated e.g. via another NN \cite{hunt1991neural, rivals2000nonlinear, bonassi2022recurrent}.
However, thanks to their control-affine structure, such an inverse is explicitly available for CA-NNARX models, and it is hence leveraged to notably simplify the IMC synthesis.}

\REV{Finally, the proposed control strategy is tested on a real, control affine laboratory apparatus, i.e., the quadruple tank process often used, in simulation, as a benchmark for control algorithms \cite{Johansson2000TheQP,  alvarado2011comparative}.
This experimental validation shows the importance of learning $\delta$ISS models to achieve reliable closed-loop performance in spite of measurement noise and plant-model mismatch.
In addition, we compare in simulation the performance of the proposed IMC scheme with that of MPC, showing that the two achieve similar tracking performance, despite the computational cost of the former being a small fraction of that of the latter.}


\subsection{Paper structure}
The paper is organized as follows. In Section \ref{sec:cannarx}, the structure of the proposed CA-NNARX model is introduced, and its $\delta$ISS  property is investigated.
In Section \ref{sec:controldesign} the IMC control scheme is described, together with its closed-loop properties.
In Section \ref{sec:simulation}, the proposed approach is tested on the Quadruple tank system and the closed-loop performance of IMC are discussed.
Lastly, some conclusions are drawn in Section \ref{sec:conclusion}. The proofs of the theoretical results are reported in the Appendix to ease the readability of the paper.
\subsection{Notation}
Given a vector $v$,  $v^\prime$ is its transpose and $[v]_i$ its $i$-th element.
Vectors' $p$-norms are denoted by $\|v\|_p$.
The Hadamard, or element-wise, product between two vectors $u$ and $v$ is denoted by $u\otimes v$.
The discrete time index is indicated as a subscript, i.e., $v_k$ indicates vector $v$ at time $k \in \mathbb{Z}_{\geq 0}$.
Sequences of time vectors are represented by boldface fonts, for example
$\textbf{v} = \{v_0, v_1, ...\}$.
Note that we denote  $\|\textbf{v}\|_{\infty} = \max_{k\geq }\|v_k\|$.
 We indicate by $I_{n, m}$ and $0_{n,m}$ the $n$-by-$m$ identity and the $n$-by-$m$ null matrices, respectively.
 We define $\diag(A_1, ..., A_n)$ the block-diagonal operator, i.e. the operator yielding a matrix having the blocks $A_1, ..., A_n$ on its diagonal.
{For a set $\mathcal{X}$, we let $\Int(\mathcal{X})$ be the interior of $\mathcal{X}$.}

\section{Control Affine NNARX Models}\label{sec:cannarx}
\subsection{Model structure}\label{sec:controlaffnnarx}
Consistently with NNARX models~\cite{bonassi2021nnarxstability}, in CA-NNARX models we let the output $y \in \mathcal{Y} \subset \mathbb{R}^{n_y}$ at time $k+1$ be a nonlinear regression over $H$ past outputs $y$ and inputs $u \in \mathcal{U}\subset \mathbb{R}^{n_u}$ up to time $k$, i.e.
\begin{equation}\label{eq:model:dynamics}
\REV{\hat{y}}_{k+1} = \eta(y_k, y_{k-1}, ..., y_{k-H+1}, u_k, u_{k-1}, ..., u_{k-H}; \bm{\Phi})
\end{equation}
In \eqref{eq:model:dynamics}, we let $\eta$ be a nonlinear regression function, parametrized by $\bm{\Phi}$, that is affine with respect to the control variable $u_k$. 
The specific structure of this regression function is described later in this chapter.
At this stage, let us define
\begin{equation} \label{eq:model:nnarx_states}
	z_{h, k} = \left[\begin{array}{c}
		y_{k-H+h} \\
		u_{k-H+h-1}
	\end{array}\right]
\end{equation}
with $z_{h, k}\in \mathbb{R}^{n_z}$, $n_z = n_u + n_y$ and $h \in \{1, ..., H \}$.
We can hence recast model \eqref{eq:model:dynamics} in a state-space normal form, which reads as
\begin{equation} \label{eq:model:nnarx_normal_form}
\begin{dcases}
		z_{1, k+1} = z_{2, k} \\
		\quad \vdots \\
		z_{H-1,k+1} = z_{H, k} \\
		z_{H, k+1} = \begin{bmatrix}
			{\eta}(z_{1, k}, z_{2, k}, ..., z_{H, k}, u_{k}; \bm{\Phi}) \\
			u_{k}
		\end{bmatrix} \\
	\REV{\hat{y}}_{k} = [I\quad 0] \, z_{H, k}
\end{dcases}
\end{equation}
Letting
$x_{k} = [ z_{1, k}^\prime, ..., z_{H, k}^\prime]^\prime \in \mathcal{X} \subseteq \mathbb{R}^{n}$ be the state vector,
\eqref{eq:model:nnarx_normal_form} can be compactly formulated as
\begin{subequations} \label{eq:model:statespace}
\begin{equation}
\begin{dcases}
  x_{k+1} = {A} x_{k} + B_{u} u_{k} + B_{x} \eta(x_{k}, u_{k}; \bm{\Phi}) \\
  \REV{\hat{y}}_{k} = C x_{k}.
\end{dcases}
\end{equation}
The system matrices $A$, $B_u$, $B_x$, and $C$ are fixed by the structure of the model, and they  read
\begin{equation} \label{eq:model:matrices}
	\begin{aligned}
		A &= {\begin{bmatrix}
		0_{n_z,n_z} & I_{n_z,n_z} & 0_{n_z,n_z} & ... & 0_{n_z,n_z} \\
		0_{n_z,n_z} & 0_{n_z,n_z} & I_{n_z,n_z} & ... & 0_{n_z,n_z} \\
		\vdots &&& \ddots & \vdots \\
		0_{n_z,n_z} & 0_{n_z,n_z} & 0_{n_z,n_z} & ... & I_{n_z,n_z} \\
		0_{n_z,n_z} & 0_{n_z,n_z} & 0_{n_z,n_z} & ... & 0_{n_z,n_z}
		\end{bmatrix}} \\
		B_u &={\begin{bmatrix}
			0_{n_z,n_u} \\
			0_{n_z,n_u} \\
			\vdots \\
			0_{n_z,n_u} \\
			\tilde{B}_u
		\end{bmatrix}}\quad
		B_x={\begin{bmatrix}
			0_{n_z, n_y} \\
			0_{n_z, n_y} \\
			\vdots \\
			0_{n_z, n_y} \\
			\tilde{B}_x
		\end{bmatrix}} \\
		C &= \begin{bmatrix}
			0_{n_y, n_z} & ... & 0_{n_y, n_z} & \tilde{C}
	\end{bmatrix}
	\end{aligned}
\end{equation}
where the blocks $\tilde{B}_u$, $\tilde{B}_{x}$, and $\tilde{C}$ are defined as follows:
\begin{equation}
	\tilde{B}_u = \begin{bmatrix}
		0_{n_y, n_u}\\
		I_{n_u, n_u}
	\end{bmatrix}, \,\,\,
	\tilde{B}_x = \begin{bmatrix}
		I_{n_y, n_y} \\
		0_{n_u, n_y}
	\end{bmatrix}, \,\,\,
	\tilde{C} = \begin{bmatrix}
		I_{n_y, n_y} & 0_{n_y, n_u}
	\end{bmatrix}
\end{equation}
\end{subequations}

\begin{figure}[t]
	\centering	\includegraphics[width=1\columnwidth]{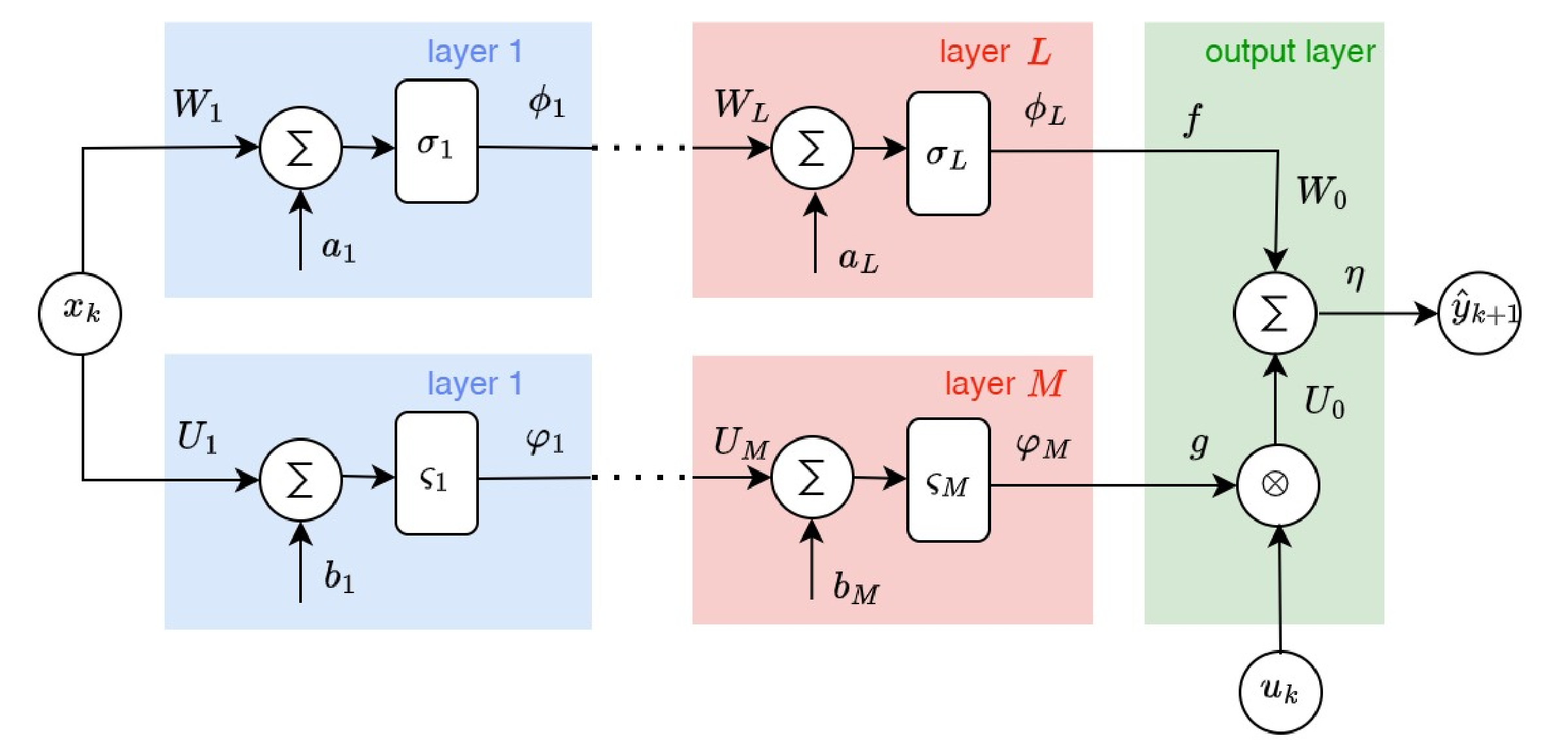}
	\caption{Structure of the proposed CA-NNARX model.}
	\label{fig:cannarx}
\end{figure}

{Concerning the regression function, since the structure of the model is assumed to match the control-affine structure of the system, we let $\eta$ be a control-affine feedforward neural network, as sketched in Figure \ref{fig:cannarx}.}
More specifically, this network takes the structure
\begin{equation}\label{eq:model:controlaffine_nn}
    \eta(x_k, u_k; \bm{\Phi}) = W_0f(x_k) + U_0\left (g(x_k) \otimes u_k \right )
\end{equation}
where $W_0$  and $U_0$ are learnable weight matrices of proper dimensions, while $f : \mathbb{R}^n \rightarrow \mathbb{R}^v $ and $g : \mathbb{R}^n \rightarrow \mathbb{R}^{n_u}$ are feed-forward neural networks with $L$ and $M$ layers, respectively. Note that $x_k$ is independent on $u_k$.
As a feed-forward NN, the function $f(x_k)$ is  defined as the concatenation of $L$ nonlinear transformations,
\begin{subequations}\label{eq:model:ffnn_f}
\begin{equation}
	f(x_k) = (\phi_L \circ ... \circ \phi_1) (x_k)
\end{equation}
where $\circ$ denotes functions' composition and $\phi_i$, with $i \in \{ 1, ..., L \}$, represent the nonlinear transformation applied by the $i$-th layer. That is,
\begin{equation}
	\phi_i(v) = \sigma_i(W_i v + a_i)
\end{equation}
with $W_i$ and $a_i$ being the weight matrix and bias of the layer, and $\sigma_i(\cdot)$ its nonlinear activation function applied element-wise on its vector argument.
We here assume that $\sigma_i(\cdot)$ is Lipschitz-continuous with Lipschitz constant $\Lambda_i$, and zero-centered, meaning that $\sigma_i(0) = 0$.
\end{subequations}
\begin{subequations}\label{eq:model:ffnn_g}
Similarly, function $g(x_k)$ is defined as the concatenation of $M$ nonlinear transformations,
\begin{equation}
	g(x_k) = (\varphi_M \circ ... \circ \varphi_1) (x_k)
\end{equation}
where each layer $j \in \{ 1, ..., M \}$ is defined as
\begin{equation} \label{eq:model:ffnn_g:layer}
	\varphi_j(v) = \varsigma_j(U_j v + b_j)
\end{equation}
In \eqref{eq:model:ffnn_g:layer}, $U_j$ and $b_j$ denote the weight matrix and bias vector of the $j-th$ layer.
The nonlinear function $\varsigma_j(\cdot)$ is assumed to have a Lipschitz constant $\tilde{\Lambda}_j$ and to be radially bounded, i.e., $\varsigma_j(\cdot) \in (\ubar{\varsigma}_i, \bar{\varsigma}_i)$. For the sake of clarity, we assume that $(\ubar{\varsigma}_i, \bar{\varsigma}_i) \subseteq (-1,  1)$. Note that both the $\tanh$ and sigmoid activation functions satisfy this condition.
\end{subequations}
The set of trainable weights is therefore defined as
\begin{equation}
\bm{\Phi} = \left \{ W_0, U_0, \left \{W_i, a_i\right \}_{i \in \{ 1,...,L\}}, \left \{U_j, b_j\right \}_{j \in \{ 1,...,M \} } \right \}
\end{equation}

The CA-NNARX model is hence described by the NARX state-space model \eqref{eq:model:statespace} combined with the control-affine nonlinear regressor \eqref{eq:model:controlaffine_nn}.
In what follows, let us compactly denote the resulting model as
\begin{equation}\label{eq:model:compact_model}
	\Sigma(\bm{\Phi}): \, \begin{dcases}
 	 	x_{k+1} = F(x_k, u_k; \bm{\Phi}) \\
 	 	\REV{\hat{y}}_{k} = G(x_{k}; \bm{\Phi})
	\end{dcases}
\end{equation}
where functions $F(x_k, u_k; \bm{\Phi})$ and $G(x_{k}; \bm{\Phi})$ can be straightforwardly derived from \eqref{eq:model:statespace} and \eqref{eq:model:controlaffine_nn}.

\subsection{Stability of CA-NNARX models}
We can now investigate the stability properties of the proposed CA-NNARX architecture, with the aim of providing a sufficient condition under which the $\delta$ISS of this class of models can be ensured.
To this end, let us first assume the boundedness of the input set with respect to which the stability is investigated.
\begin{assumption} \label{ass:input}
	The input vector $u_k$ is unity-bounded, i.e. $$u_k \in \mathcal{U} = [-1, 1]^{n_u}.$$
\end{assumption}
Note that Assumption \ref{ass:input} is not restrictive since, as long as the input is bounded in a finite set, it can always be satisfied via suitable normalization procedures \cite{bonassi2022recurrent}.

\smallskip
We now recall a couple of definitions that are required to state the $\delta$ISS property.
\begin{definition}[$\mathcal{K}_\infty$ function]
	Function $\gamma(s): \mathbb{R}_{\geq 0} \rightarrow \mathbb{R}_{\geq 0}$ is of class $\mathcal{K}_\infty$ if $\gamma(0) = 0$, it is strictly increasing, and $\gamma(s) \to +\infty$ when $s \to +\infty$.
\end{definition}
\begin{definition}[$\mathcal{KL}$ function]
	Function $\beta(s, t): \mathbb{R}_{\geq 0} \times \mathbb{R}_{\geq 0}  \rightarrow \mathbb{R}_{\geq 0}$ is of class $\mathcal{KL}$ if it is of class $\mathcal{K}_\infty$ with respect to its first argument and, for any $s \in \mathbb{R}_{\geq 0}$, satisfies $\beta(s, t) \to 0$ when $t \to +\infty$.
\end{definition}
In light of these definitions, the stability notion here considered is defined as follows.
\begin{definition}[$\delta$ISS \cite{bayer2013discrete}] \label{def:deltaiss}
	System \eqref{eq:model:compact_model} is Incrementally Input-to-State Stable ($\delta$ISS) if there exist functions $\beta \in \mathcal{KL}$ and $\gamma \in \mathcal{K}_\infty$ such that, for any pair of initial states $x_{a, 0} \in \mathcal{X}$ and $x_{b, 0}  \in \mathcal{X}$, any pair of input sequences $\bm{u}_a = \{ u_{a,0}, ..., u_{a, k-1} \} $ and $\bm{u}_b = \{ u_{b,0}, ..., u_{b, k-1} \} $ (where $u_{a, t} \in \mathcal{U}$ and $u_{b,t}\in \mathcal{U}$ for all $t$), and any discrete-time step $k \in \mathbb{Z}_{\geq 0}$, it holds that
	\begin{equation} \label{eq:deltaiss:definition}
	\scalemath{0.825}{
		\| x_{a,k} - x_{b,k} \|_2 \leq \beta(\| x_{a,0} - x_{b,0} \|_2, k) + \gamma \left(\max_{\tau \in \{ 0, ..., k-1 \} }\| u_{a,\tau} - u_{b,\tau} \|_{2} \right)}
	\end{equation}
	where $x_{\alpha,k}$ denotes the state trajectory of the system initialized in $\bar{x}_{\alpha}$ and fed by the sequence $\bm{u}_{\alpha}$, with $\alpha \in \{a, b\}$.
\end{definition}

Therefore, the $\delta$ISS property implies that the smaller the distance between two input trajectories, the tighter the asymptotic bound on the resulting state trajectories. 
\REV{It has been observed~\cite{bonassi2023reconciling} that the $\delta$ISS property implies a degree of robustness to input perturbations by the model. As shown in Section~\ref{sec:simulation}, model's robustness allows for model-based controllers with closed-loop performance that are less sensitive to e.g. measurement noise.}
Another reason why $\delta$ISS is particularly useful is that it implies  (see \cite{bayer2013discrete} and \cite{bonassi2023reconciling}) the weaker, and well-known, ISS property \cite{jiang2001input}, which entails that bounded input sequences yield bounded state (and output) trajectories.
Specifically, with the aim of synthesizing the IMC scheme, we are interested in the following input-output stability property.

\begin{definition}[IOS] \label{def:ios}
	System \eqref{eq:model:compact_model} is Input-to-Output Stable (IOS) if there exist functions $\tilde{\beta}_y \in \mathcal{KL}$ and $\tilde{\gamma}_y \in \mathcal{K}_\infty$, and a scalar $\varrho_y \geq 0$, such that, for any initial state $x_0 \in \mathcal{X}$, any input sequence $\bm{u} = \{ u_0, ..., u_{k-1} \}$ (where $u_t \in \mathcal{U}$ for all $t$), and any time-step $k \in \mathbb{Z}_{\geq 0}$, it holds that
	\begin{equation} \label{eq:iops:definition}
		\| \REV{\hat{y}}_{k} \|_2 \leq \tilde{\beta}_y(\| x_{0} \|_2, k) + \tilde{\gamma}_y \left(\max_{\tau \in \{ 0, ..., k-1 \} }\| u_{\tau} \|_2 \right) + \varrho_y
	\end{equation}
	where $\REV{\hat{y}}_k$ denotes the output of system \eqref{eq:model:compact_model} when it is initialized in $x_0$ and it is fed by the input sequence $\bm{u}$.
\end{definition}
\begin{proposition}[$\delta$ISS implies IOS, \cite{bonassi2023reconciling}] \label{prop:deltaiss_implies_ios}
	If system \eqref{eq:model:compact_model} is $\delta$ISS and its equilibrium manifold is non-empty, then it is also IOS.
\end{proposition}
The following novel result concerning the $\delta$ISS and hence the IOS of CA-NNARX models can now be stated.
\begin{theorem} \label{thmdISS}
A sufficient condition for the $\delta$ISS of the CA-NARX model is that
\begin{equation}\label{eq:delta_iss_condition}
\| W_0 \|_2 \, \prod_{i=1}^{L} \Lambda_i \norm{W_i}_2 + \| U_0 \|_2 \, \prod_{j=1}^{M} \tilde{\Lambda}_j \norm{U_j}_2 \leq \frac{1}{\sqrt{H}}
\end{equation}
\end{theorem}
\begin{proof} \nonumber
	See Appendix \ref{app1}.
\end{proof}

Notably, the sufficient condition \eqref{eq:delta_iss_condition}  can be used (\emph{i}) \emph{a posteriori}, to certify the $\delta$ISS and IOS properties of a given CA-NNARX model, or (\emph{ii}) \emph{a priori}, to enforce the $\delta$ISS and IOS during the training procedure itself, so as to learn provenly-stable CA-NNARX models.
As discussed in Section~\ref{sec:simulation}, this second approach is followed here by including \eqref{eq:delta_iss_condition} as a (soft) constraint in the training procedure.
\REV{This form of regularization allows training to be steered to regions in the space of the weights where models are guaranteed to be stable and robust to perturbations on input signals.}

\REVV{Finally, it should be pointed out that the condition proposed in Theorem~\ref{thmdISS} is only sufficient and potentially conservative, meaning that the model could display state trajectories compatible with the $\delta$ISS property even though this property cannot be guaranteed by Theorem~\ref{thmdISS}.}

\section{Internal Model Control design} \label{sec:controldesign}
Let us, at this stage, assume that a CA-NNARX model $\Sigma(\bm{\Phi}^\star)$ of the system, see \eqref{eq:model:compact_model}, has been trained, and that it satisfies the stability condition \eqref{eq:delta_iss_condition}.
The goal is now to synthesize an IMC architecture that allows to track a piecewise-constant reference signal $y^o$.
To this end, let us first take the following customary assumption.
\begin{assumption}
	The setpoint $y^{o}$ is feasible, i.e., there exist $u^o \in  \Int(\mathcal{U})$ and $x^o \in \Int(\mathcal{X})$ such that $(x^o, u^o, y^o)$ is an equilibrium of system \eqref{eq:model:compact_model}.
\end{assumption}
\REV{Note that Assumption 2 could in principle be relaxed by designing a reference governor \cite{gilbert1995discrete} but we will, for simplicity, assume here that the assumption holds.}
In the following we briefly summarize the IMC approach, and we then show how it specializes to CA-NNARX models.

\subsection{Internal Model Control structure and ideal properties}\label{sec:imc}

\begin{figure}[b]
	\centering
	\includegraphics[width=0.9\columnwidth]{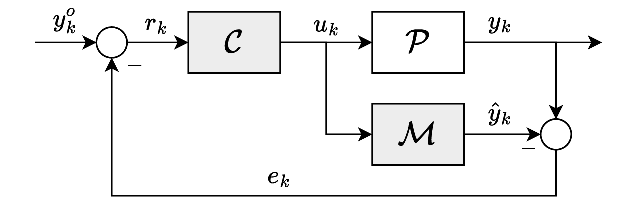}
	\caption{General scheme of Internal Model Control with reference tracking.}
	\label{fig:IMCscheme}
\end{figure}

Having introduced the class of models used to learn the system from  data, and having shown how to impose its stability and robustness properties, we can now introduce the IMC scheme.
\REV{Underlying the IMC control scheme is the diagram shown in Figure~\ref{fig:IMCscheme}, which consists of three blocks}: the  (unknown) plant $\mathcal{P}$, the (learned) model $\mathcal{M}$ of this plant, and the controller $\mathcal{C}$.
The rationale of this control strategy is simple: if the model is perfect ($\mathcal{P}=\mathcal{M}$), the feedback connection is open (\REV{$e_{k} = y_{k} - \hat{y}_{k} = 0$}) so that, selecting the controller as the inverse of the model ($\mathcal{C}=\mathcal{M}^{-1}$), perfect control is achieved, i.e., $y_{k} = \hat{y}_{k} = r_k = y^o_k$.
\REVV{If instead there is any plant-model mismatch, the modeling error signal $e_k$ is fed back, allowing the controller to compensate for it.}

\smallskip
The IMC strategy enjoys the following well-known ideal properties \cite{economou1986internal}, reported below.

\begin{property}(Stability \cite{economou1986internal})
If the model is perfect, i.e., $\mathcal{M} = \mathcal{P}$, and both $\mathcal{C}$ and $\mathcal{P}$ are input-output stable, then the overall system is input-output stable.
\end{property}

\begin{property}(Perfect control \cite{economou1986internal})
Assume the model is perfect, that the controller matches the model's inverse, i.e.  $\mathcal{C} = \mathcal{M}^{-1}$, and that both blocks are input-output stable.
Then, perfect control is achieved, i.e., $y_{k} = y^o_k$.
\end{property}

\begin{property}(Zero offset \cite{economou1986internal})
Assume that the model is perfect, and that the steady state control action generated by the controller matches the steady-state value of the model’s inverse.
Then, if both blocks are input-output stable, offset-free tracking is asymptotically attained.
\end{property}

It is worth noticing that these properties are ideal, because they hold as long as the model and the controller are initialized to specific initial states, see \cite[Chapter 8]{bonassi2023reconciling}.
Finding these correct initial conditions might not be possible in many applications, e.g., due to measurement noise.
However, when the plant, the model, and the controller are $\delta$ISS, the requirement of exact initialization can be dropped while still achieving these properties asymptotically, see again \cite{bonassi2023reconciling}.
\REV{This is indeed one of the reasons why enforcing the $\delta$ISS of the IMC blocks is recommendable.}

\REV{One of the main problems of IMC is that this scheme calls, in principle, for} the exact model inverse.
\REV{This is especially challenging when the model $\mathcal{M}$ is a black-box RNN model identified from the data, since} not only retrieving an analytical inverse of such a model might not be possible, but an inverse might not even exist.
\REV{To obviate this problem,}  in \cite{bonassi2022imc} the \REV{approximation capabilities of RNNs are exploited to approximate the model inverse with a suitably trained $\delta$ISS RNN.}
As shown in the following section, the structure of CA-NNARX models can instead be leveraged to retrieve the controller explicitly.

\begin{figure}[t]
	\centering
	\includegraphics[width=1\columnwidth]{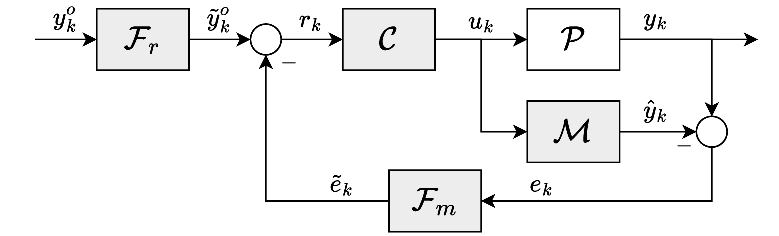}
	\caption{Internal model control architecture with model reference and low-pass filter on the feedback.}
	\label{fig:IMCscheme2}
\end{figure}

\REV{Finally, it must be noted that variants to the basic IMC scheme have been proposed to encode the desired closed-loop performance, as well as to enhance the robustness of the closed-loop \cite{Morari1989robust, economou1986internal}.
The first objective is achieved by pre-filtering the reference signal $y^o_k$ with the model reference $\mathcal{F}_r$, which encodes the  desired closed-loop dynamics.
The second objective is instead achieved by filtering the modeling error feedback $e_{k}$ with a low-pass filter $\mathcal{F}_m$ to cut the high-frequency measurement noise on the output $y_k$.
The resulting IMC scheme is depicted in Figure \ref{fig:IMCscheme2}.}

\begin{remark} \label{rem:imcdesign}
\REV{
	The blocks $\mathcal{F}_r$ and $\mathcal{F}_m$ are, for the sake of simplicity, usually taken equal ($\mathcal{F}_r = \mathcal{F}_m$), c.f. \cite{economou1986internal}, and  typically chosen as low-pass filters with unitary static gain and with the same time constant. A standard design procedure is to choose their time constant to be roughly equivalent to the one desired for the closed-loop system. 
	In this setting, this time constant is the only free design parameter of the IMC scheme.}
\end{remark}

\vspace{-2mm}
\begin{remark}\REV{
This IMC control strategy is sometimes termed Inverse Model Control \cite{zhai2023model}, especially in the context of models learned through fuzzy networks \cite{kumbasar2017inverse}, stressing the fact that the controller is designed to approximate the model inverse.
However, in view of its long-term popularity, we will use the terminology \emph{Internal Model Control} in this article,  given that the idea of modeling the process inverse has been successfully exploited in other synthesis procedures \cite{dirion1995design}.}
\end{remark}

\subsection{Internal Model Control for CA-NNARX models} \label{sec:cannarx_imc}

The structure of CA-NNARX models and the measurability of their state vector\footnote{{The state $x_k$ is composed of past inputs and outputs, and is thus known.}} allow to define the model's inverse explicitly.
Indeed, recalling \eqref{eq:model:dynamics} and \eqref{eq:model:controlaffine_nn}, the state-dependent input-output relationship reads as
\begin{equation}\label{eq:sysmodel}
    \hat{y}_{k+1} = W_0 f(x_k) + U_0( g(x_k) \otimes u_k )
\end{equation}
where $f(x_k)$ and $g(x_k)$ are the feed-forward NNs defined in \eqref{eq:model:ffnn_f} and \eqref{eq:model:ffnn_g}, respectively.
Let us then consider the following assumption.
\begin{assumption}\label{ass:boundedG}
	For any state $x \in \mathcal{X}$, any component of $g(x)$ is strictly different from zero.
	That is, there exists $\epsilon > 0$ such that, for any $j \in \{ 1, ..., n_u\}$,
	\begin{equation}\label{eq:imc:g_ass}
       \inf_{x \in \mathcal{X} } \,\, \big\lvert \big[ g(x) \big]_j \big\lvert \geq \epsilon
	\end{equation}
\end{assumption}
\begin{remark}
   \REV{The fulfillment of Assumption \ref{ass:boundedG} can either be guaranteed \emph{a posteriori}, i.e., after the model's training procedure by solving an optimization problem seeking to maximize $\epsilon$ numerically (see Section \ref{sec:simulation}), or \emph{a priori} by properly structuring the feed-forward neural network $g(x)$.
    In particular, by adopting a sigmoidal activation function for the last layer $\big(\varsigma_M(x) = \frac{1}{1 + e^{-x}} \in (0, 1) \big)$, one can guarantee that Assumption \ref{ass:boundedG} is fulfilled.}
\end{remark}

At this stage, under the assumption of constant, \REV{or slowly time varying,} reference signal (\REV{$\tilde{y}^o_{k+1} = \tilde{y}^o_k$}), one can define the control action based on the  nominal case as
\begin{equation}\label{eq:imc:inverse}
    u_{k} = \Sat_{\,{\mathcal{U}}} \Big( \REV{\diag(g(x_k))^{-1}} \big[ U_0^{\dag}( r_{k} - W_0 f(x_k)) \big]  \Big)
\end{equation}
where \REV{$r_k = \tilde{y}^o_k - \tilde{e}_{k}$} denotes the reference signal adjusted by the modeling error feedback, and $U_0^\dag$ denotes the Moore-Penrose inverse of $U_0$.
\REV{Note that inequality \eqref{eq:imc:g_ass} and the definition of $U_0^\dagger$ allow \eqref{eq:imc:inverse} to be always well-defined.
}
{What is more, $\Sat_{\,{\mathcal{U}}}(v)$ indicates the projection of the argument into the input set ${\mathcal{U}}$}. This guarantees the input constraint satisfaction and ensures the fulfillment of Assumption \ref{ass:input}.

Note that, owing to the continuity of the system's functions $f(\cdot)$ and $g(\cdot)$, the saturated control law \eqref{eq:imc:inverse} matches the model's inverse in a neighborhood of the equilibrium, as shown in the following proposition.

\begin{proposition}
For any target equilibrium $(x^o, u^o, y^o)$ there exists a neighborhood $\Gamma(x^o) \subseteq \mathcal{X}$ such that, for any $x_k \in \Gamma(x^o)$, the control law \eqref{eq:imc:inverse} matches the nominal model inverse.
\end{proposition}
\begin{proof}
	First, let us recall that the functions $f(x_k)$ and $g(x_k)$ are Lipschitz continuous and that, in view of Assumption \ref{ass:boundedG}, the pair $(u^o, x^o)$ solves \eqref{eq:imc:inverse}, where $r^o = y^o$ as the nominal model is considered.
	By continuity arguments, there exists a sufficiently small neighborhood $\Gamma(x^o)$ where, $\forall x \in \Gamma(x^o)$,
	\begin{equation*}
		\Big\| \REV{\diag(g(x_k))^{-1}}  \big[ U_0^{\dag}( r^o - W_0 f(x)) \big] \Big\|_\infty < 1
	\end{equation*}
	so that the saturation in \eqref{eq:imc:inverse} is inactive.
	The exact model inverse is therefore recovered.
\end{proof}

We are now in the position of discussing closed-loop stability properties. To this end, let us assume that the plant dynamics are described by the model's equations, \eqref{eq:model:compact_model}, with some (bounded) additive output disturbance $d_k \in \mathcal{D}$, i.e.
\begin{equation} \label{eq:imc:plant}
	\mathcal{P}: \begin{dcases}
		\chi_{k+1} = F(\chi_k, u_k; \bm{\Phi}^\star) \\
		y_{k} = G(\chi_k; \bm{\Phi}^\star) + d_k
	\end{dcases}
\end{equation}
Then, the following closed-loop property holds.

\begin{proposition} \label{prop:iostability}
	If the CA-NNARX model satisfies Theorem \ref{thmdISS} and the plant is described by \eqref{eq:imc:plant}, then the closed-loop is input-output stable under the control law \eqref{eq:imc:inverse}.
\end{proposition}
\begin{proof}
First, let us notice that the saturation included in the control law \eqref{eq:imc:inverse} allows to fulfill Assumption \ref{ass:input}, meaning that the model's $\delta$ISS is ensured.
By Proposition \ref{prop:deltaiss_implies_ios}, the model is also IOS.
Because $\mathcal{D}$ is bounded this, in turn, implies that the plant \eqref{eq:imc:plant} is IOS. The modeling error feedback $\tilde{e}_{k}$ is thus also bounded.
Therefore, for any finite reference $y^o_k$ and modeling error feedback, the control action is bounded in $\mathcal{U}$, and for any such control variable the IOS property guarantees the boundedness of the plant's output and the closed loop is input-output stable.
\end{proof}
%

\section{Numerical example}\label{sec:simulation}
\subsection{Benchmark system}
\begin{figure}[t]
\centering
    \includegraphics[width=0.55\columnwidth]{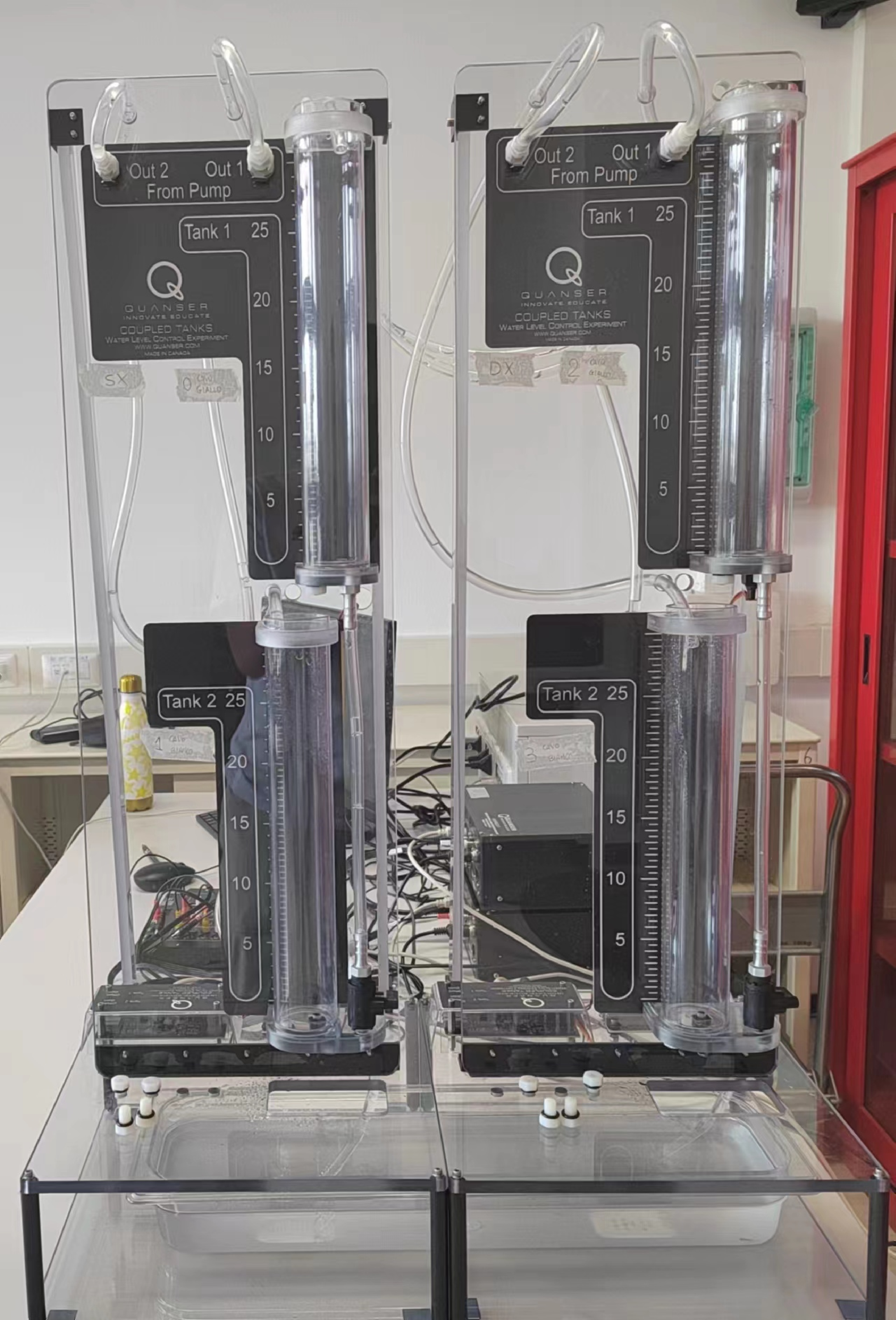}
	\caption{Quadruple Tank lab apparatus used for the closed-loop validation of the proposed approach.}
    \vspace{4mm}
	\label{fig:4tank}
	\centering	\includegraphics[width=0.8\columnwidth]{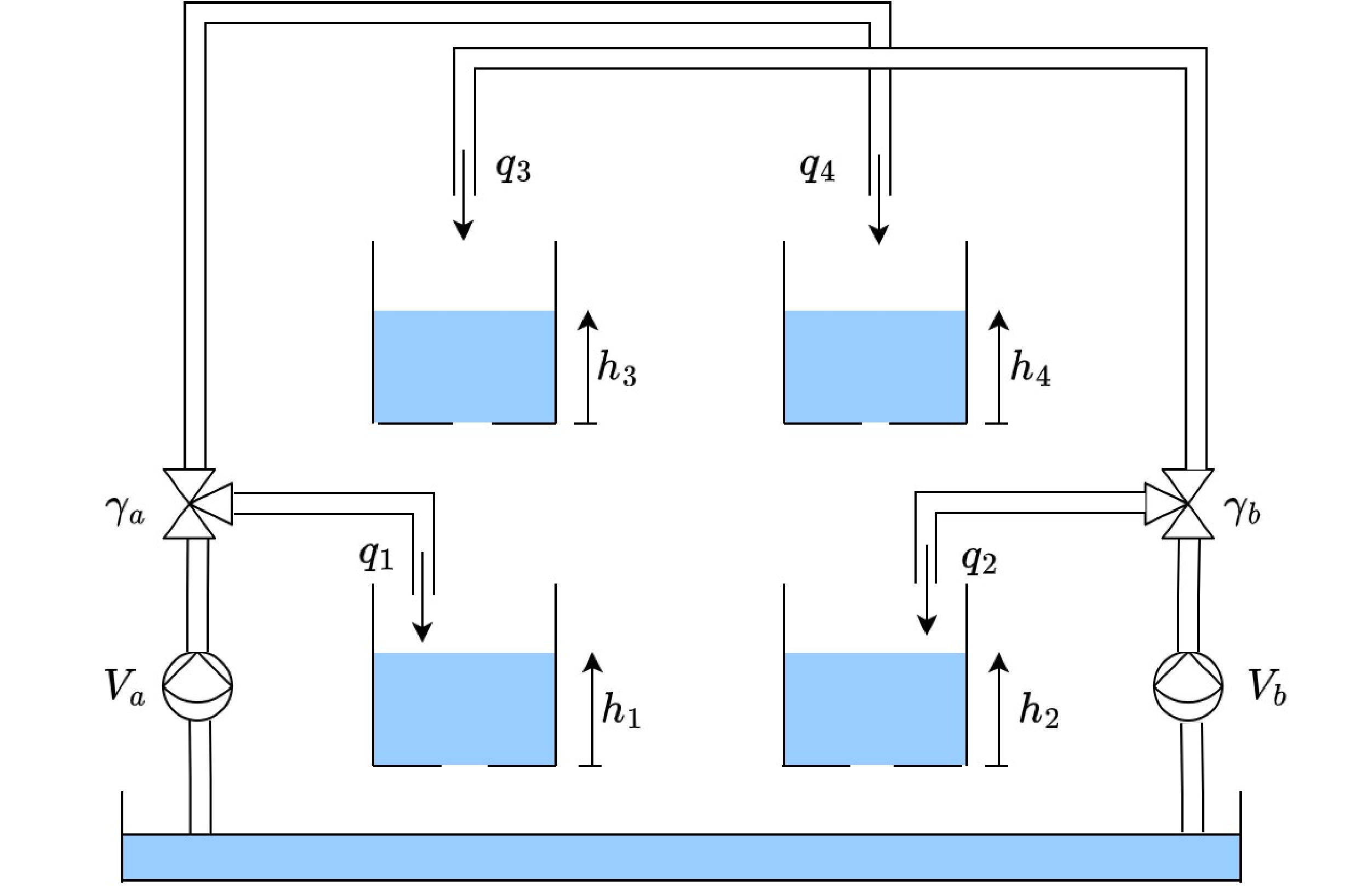}
	\caption{Schematic of Quadruple Tank system}
	\label{fig:Benchmark}
\end{figure}

The proposed scheme has been tested on the Quadruple Tank benchmark reported in \cite{Johansson2000TheQP}, \cite{bonassi2022imc}, and used in \cite{alvarado2011comparative} to compare the behavior of different distributed MPC algorithms.
 An illustration of the Quadruple Tanks laboratory apparatus \REV{used for the closed-loop validation experiments} is depicted in Figure \ref{fig:4tank}, whereas the corresponding abstract system is depicted in Figure \ref{fig:Benchmark}. T
 he system consists of four water tanks with levels $h_1$, $h_2$, $h_3$ and $h_4$, which represent the measured outputs of the system. 
The water levels are controlled by two pumps, that provide the water flows $q_a$ and $q_b$.
\REV{These flow rates depend on the applied voltages $V_a$ and $V_b$, which are the manipulable variables.}
The water flow $q_a$ is split into $q_1$ and $q_4$, where $q_1 = \gamma_aq_a$ and $q_4 = (1-\gamma_a)q_b$. Similarly, $q_b$ is split into $q_2$ and $q_3$, where $q_2 = \gamma_bq_b$ and $q_3 = (1-\gamma_b)q_b$. 
The dynamics of the system are described by
\begin{equation}\label{serbatoi}
\begin{aligned}
    \dot{h}_{1} &= - \frac{a_1}{S}\sqrt{2gh_1}+\frac{a_3}{S}\sqrt{2gh_3}+\frac{\gamma_a}{S}q_a\\
    \dot{h}_{2} &= - \frac{a_2}{S}\sqrt{2gh_2}+\frac{a_4}{S}\sqrt{2gh_4}+\frac{\gamma_b}{S}q_b\\
    \dot{h}_{3} &= - \frac{a_3}{S}\sqrt{2gh_3}+\frac{1-\gamma_b}{S}q_b\\
    \dot{h}_{4} &= - \frac{a_4}{S}\sqrt{2gh_4}+\frac{1-\gamma_a}{S}q_a\\ 
    \REV{q_a} &\REV{= \kappa_a V_a \qquad q_b = \kappa_b V_b}
\end{aligned}
\end{equation}
where the nominal values of the parameters are reported in Table \ref{tab:parameter}. 
\begin{table}[b]
	\centering
	\caption{Nominal parameters in benchmark system}
	\label{tab:parameter}
	\begin{tabular}{ccc|ccc}
		\toprule
		\textbf{Parameter}  & \textbf{Value}& \textbf{Unit} &\textbf{Parameter}  & \textbf{Value}& \textbf{Unit}  \\ \midrule
		$a_i$ & $0.1781 $&$\text{cm}^2$ &$\gamma_a$ & 0.36 &\\
		$S$ & $15.5179$&$\text{cm}^2$&$\gamma_b$ & 0.36 & \\
		$\kappa_a$ & 3.3 & $\text{cm}^3 \, V / s$  & $\kappa_b$ &  3.3 & $\text{cm}^3 \, V / s$ \\
	\bottomrule
\end{tabular}
\end{table}
Note that the input of system \eqref{serbatoi} is the pump voltage \REV{$u = [V_a, V_b]^\prime$}, whereas its output is $y = [h_1, h_2, h_3, h_4]^\prime$. 
The water levels as well as the controlled water flows are subject to saturations,
\begin{equation}\label{serbatoi_constraints}
    \begin{aligned}
        h_i &\in [0, \ 25]\ cm \quad  \forall i \in \left\{1,2,3,4\right\},\\
        V_j &\in [0, 15]\ V \quad \forall j \in \left\{a,b\right\}
    \end{aligned}
\end{equation}
{The control problem here considered is to regulate the output $y_{k}$ to the constant reference ${y}^o_k$, assumed to be a feasible equilibrium of \eqref{serbatoi}, while fulfilling \eqref{serbatoi_constraints}. 
Because system \eqref{serbatoi} is control affine, we address the control problem using the IMC strategy discussed in Section \ref{sec:cannarx_imc}. }

%

\subsection{Model identification with CA-NNARX}\label{sec:modelident}
{As we assume not to have access to the model of the system \eqref{serbatoi}, in order to synthesize the IMC law we first need to identify a CA-NNARX model from the data\REV{, following the procedure detailed in \cite[Chapter~4]{bonassi2023reconciling}}.
Hence, experiments were performed on the physical Quadruple-Tank system to obtain the training, validation, and test data. 

\smallskip
\noindent \REV{\emph{Datasets} --}
Several experiments were conducted, where the system was excited with a bivariate Multilevel Pseudo-Random Signal (MPRS) in order to explore a broad range of dynamics as well as the equilibria.  
The identification data has been sampled with sampling time $\tau_s = 1 s$, collecting a total of $21000$ points, \REV{corresponding to approximately 6 hours of operations}.}
The input and output sequences have been normalized so that they lie in $[-1, 1]$, see Assumption~\ref{ass:input}, as usual in deep learning~\cite{Good16deeplearning}.
Once the data is collected, according to the Truncated Back-Propagation Through Time (TBPTT) approach \cite{bianchi2017recurrent}, $N_{\train} = 735$ and $N_{\val} = 142$ subsequences {$\big(\bm{u}^{\{i\}}, \bm{y}^{\{i\}}\big)$} of length $T_{\train} = T_{\val} = 150$ time-steps have been extracted for the training and validation sets, respectively.
{Note that the training and validation subsequences have been extracted from independent experiments.
One experiment is used as independent test dataset, to assess the performances of the trained model. 
The test dataset consists of one sequence ($N_{\test} = 1$) of length $T_{\test} = 3000 \gg T_{\train}$.
For convenience of notation, we will denote the training, validation, and test by the sets collecting their indexes, i.e. $\mathcal{I}_{\train} = \{ 1, ..., N_{\train} \}$, $\mathcal{I}_{\val} = \{ N_{\train} + 1, ..., N_{\train} + N_{\val} \}$, and $\mathcal{I}_{\test} = \{ N_{\train} + N_{\val} + 1, ..., N_{\train} + N_{\val} + N_{\test} \}$, respectively.}

\smallskip
\noindent  \REV{\emph{Model's hyperparameters} --}
The structure of the control-affine neural network is that reported in \eqref{eq:model:controlaffine_nn}, where $f(x)$  is a two-layer feed-forward network with 15 neurons each, and $g(x)$ is a feed-forward network with 15, 15, and 2 neurons, respectively. 
Note that the number of neurons in the last layer of $g(x)$ is consistent with the number of inputs, $n_u$.
The activation function has been chosen as $\sigma = \varsigma = \text{tanh} $. {The hyperbolic tangent function is selected since it is zero-centered and has been observed to yield better performances.}
The regression horizon $H$ of the CA-NNARX model has been set to $H=5$.

\smallskip
\noindent \REV{\emph{Model training} --}
The CA-NNARX model has been identified by minimizing the simulation Mean Square Error (MSE) over the training set, i.e.,
\begin{equation}
    \bm{\Phi}^\star = \underset{\bm{\Phi}}{{\text{min}}} \, \mathcal{L}_{\bm{\text{MSE}}}(\mathcal{I}_{\train}; \bm{\Phi}).
\end{equation}

\begin{figure}[t]
    \centering
	\includegraphics[width=0.95\columnwidth]{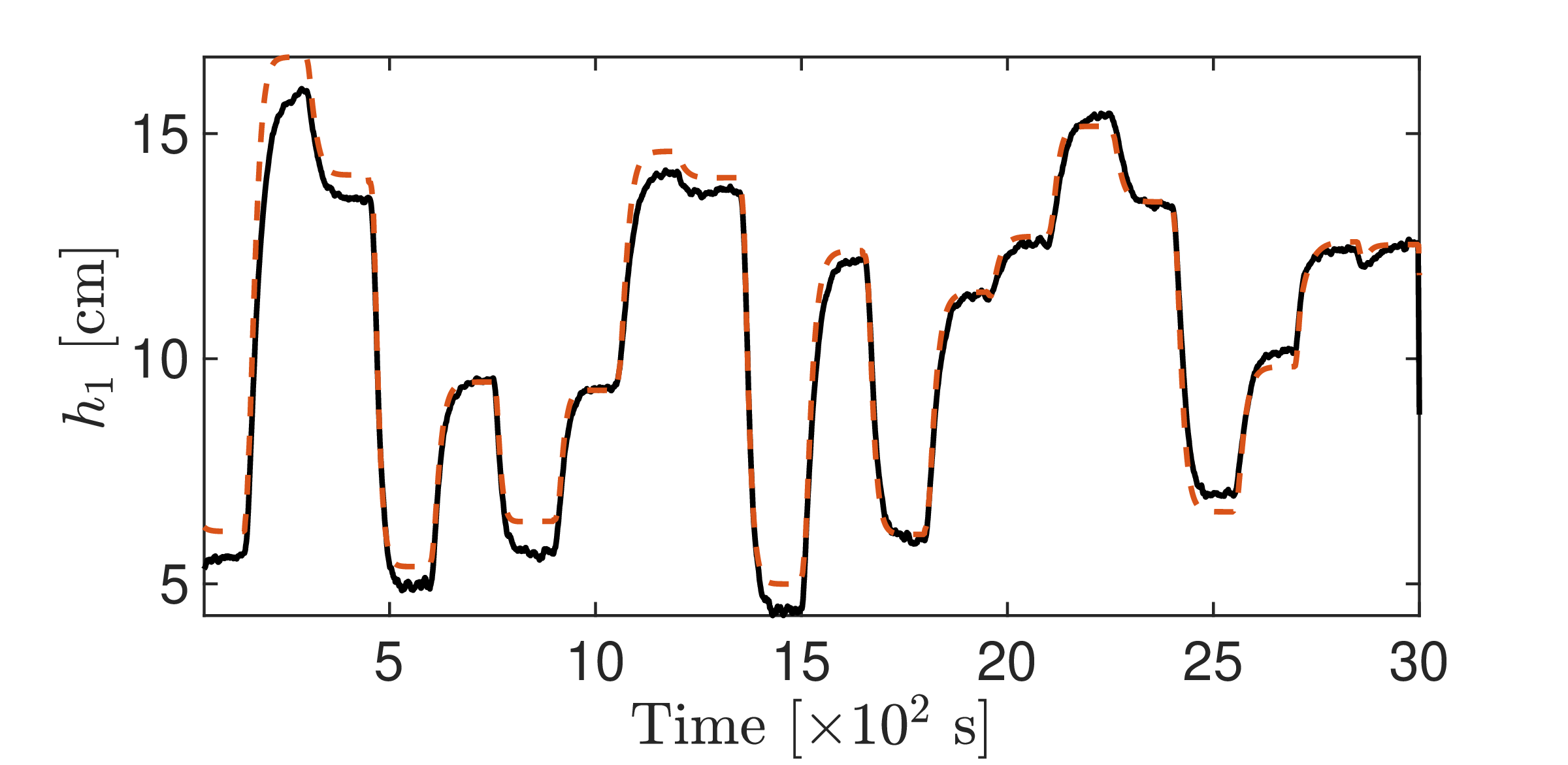}
 
    \vspace{2mm}
	\includegraphics[width=0.95\columnwidth]{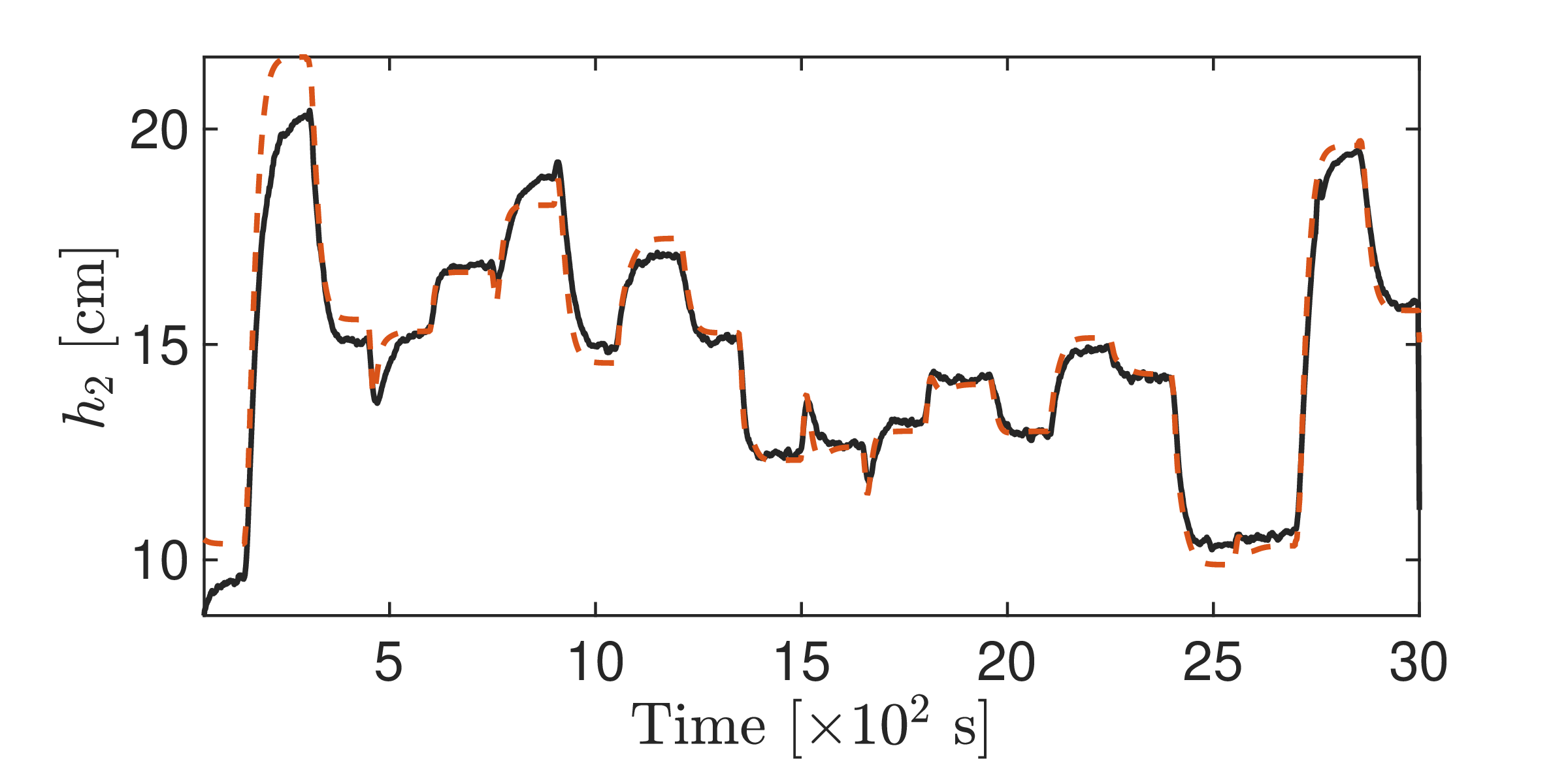}
 
    \vspace{2mm}
	\includegraphics[width=0.95\columnwidth]{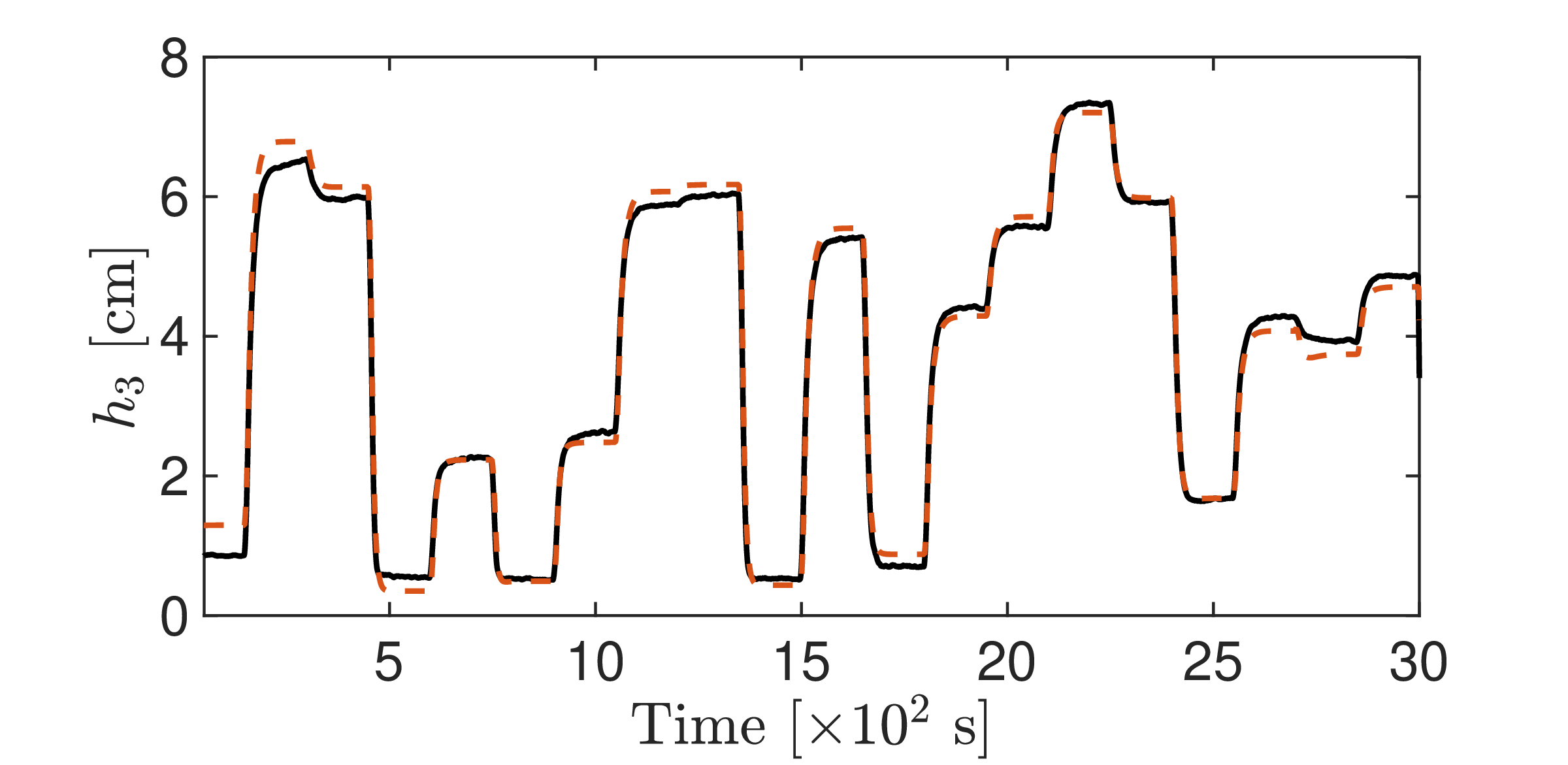}

    \vspace{2mm}
	\includegraphics[width=0.95\columnwidth]{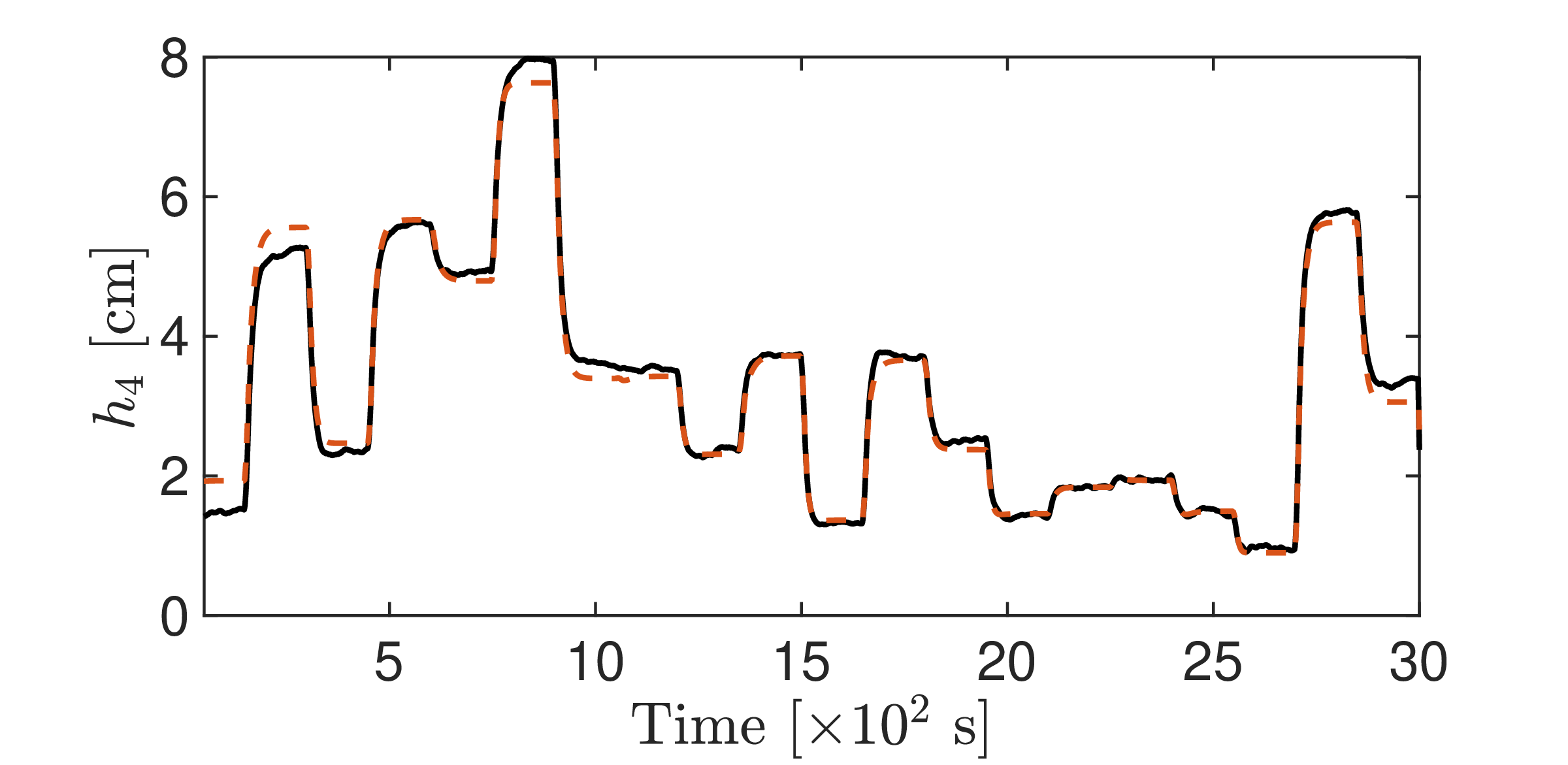}
    \caption{Open-loop prediction (red dashed line) vs ground truth (black solid line) on the test dataset of the four levels.}
    \label{fig:openlooppredition}
\end{figure}

In particular, the simulation MSE is defined as
\begin{equation}
\scalemath{0.9}{\begin{aligned}
	\mathcal{L}_{\bm{\text{MSE}}}(\mathcal{I}_{\train}; \bm{\Phi}) =& \frac{1}{\lvert \mathcal{I}_{\train} \lvert (T_{\train} - T_{w})}\sum_{i \in \mathcal{I_{\train}}} \sum_{k=T_{w}}^{T_{\train}} \Big( \hat{y}_k(x_0^{\{ i \}}, \bm{u}^{\{i\}})- y_{k}^{\{i\}} \Big)^2 \\
	&+ \rho(\nu(\bm{\Phi})),
\end{aligned}}
\end{equation}
where $\hat{y}_k(x^{\{ i\}},\bm{u}^{\{i \}})$ denotes the output of the CA-NNARX model \eqref{eq:model:compact_model} initialized in the random initial state ${x}_0^{\{ i \}}$ and fed with input sequence $\bm{u}^{\{i\}}$.
The regularization term  $\rho(\nu(\bm{\Phi}))$ is designed to enforce the $\delta$ISS property of the model.
In particular, we denote by $\nu(\bm{\Phi})$ the residual of condition \eqref{eq:delta_iss_condition}, i.e.
\begin{equation}\label{eq:regularization}
	\nu(\bm{\Phi}) = \prod_{l=0}^{L}\Lambda_l \norm{W_l}_2 + \prod_{l=0}^{M} \tilde{\Lambda}_l \norm{U_l}_2 - \frac{1}{\sqrt{H}}
\end{equation}
where the Lipschitz constant of the $\tanh$ activation function reads $\Lambda_* = \tilde{\Lambda}_* = 1$.
Noting that $\nu(\bm{\Phi}) < 0$ implies the fulfillment of condition \eqref{eq:delta_iss_condition}, the regularization term $\rho(\nu(\bm{\Phi}))$ is designed to steer the residual to slightly negative values.
As in \cite{bonassi2021grustability}, this regularization function has been selected as a piecewise linear function,
\begin{equation}
\rho(\nu(\bm{\Phi})) = \pi_{+} \left[\text{max}(\nu,-\varepsilon) +\varepsilon\right]+ \pi_{-} \left[\text{min}(\nu,-\varepsilon)+\varepsilon\right]
\end{equation}
where $\pi_{+} \gg \pi_{-} > 0$ are the weights and $\varepsilon > 0$ is a small constraint clearance.
Here, we have adopted $\pi_{-} = 10^{-4}$, $\pi_{+} = 0.035$ and $\varepsilon = 0.02$.
The model has been trained with PyTorch 1.9 for $1500$ epochs until the model's performance on the validation dataset $\mathcal{I}_{\val}$ stopped improving and the $\delta$ISS condition was fulfilled ($\nu = -0.024$).
\smallskip

\noindent \REV{\emph{Model validation ---}}
At the end of the training procedure,  the modeling performance of the trained model has been finally tested on the independent test set, \REV{where the input signal is given by a sequence of random steps within the saturation limits to avoid the overflow of the tanks' water.}
In Figure \ref{fig:openlooppredition} the model's open-loop simulation is compared to the ground truth, showing a good degree of accuracy. 
The model's accuracy has also been quantitatively evaluated by means of the $\textrm{FIT}$ index, defined as
\begin{equation}
    \text{FIT} = 100 \left( 1 -  \frac{\sum_{k=T_w}^{T_f}\big\| \hat{y}_k \big( x_{0}^{\{ i \}}, \bm{u}^{\{ i \}} \big) - {y}_{k}^{\{i\}} \big\|_2}{\sum_{k=T_w}^{T_f} \big\|  {y}_{k}^{\{i\}} - y_{\text{avg}}^{\{i\}} \big\|_2} \right)
\end{equation}
for $i \in \mathcal{I}_{\test}$, where $\hat{y}_k \big( x_{0}^{\{ i \}}, \bm{u}^{\{ i \}} \big)$ denotes the open-loop simulation of the model and $y_{\text{avg}}^{\{i\}}$ the average of $\bm{y}^{\{i\}}$.
The trained CA-NNARX model achieves a FIT value of $87.76 \%$.

\medskip
\noindent 
\emph{Comparison with a standard NNARX model}

In order to assess the advantages of the $\delta$ISS CA-NNARX architecture, two more models have been trained to serve as baselines: (\emph{i}) a standard black-box NNARX model \cite{bonassi2021nnarxstability} and (\emph{ii}) a CA-NNARX \emph{without} $\delta$ISS condition.
\REV{This allows us to assess, given the same number of learnable parameters, the advantage of imposing a control affine structure and the conservativeness of the $\delta$ISS condition provided by Theorem~\ref{thmdISS}.}

\begin{figure}[t]
    \includegraphics[width=0.95\columnwidth]{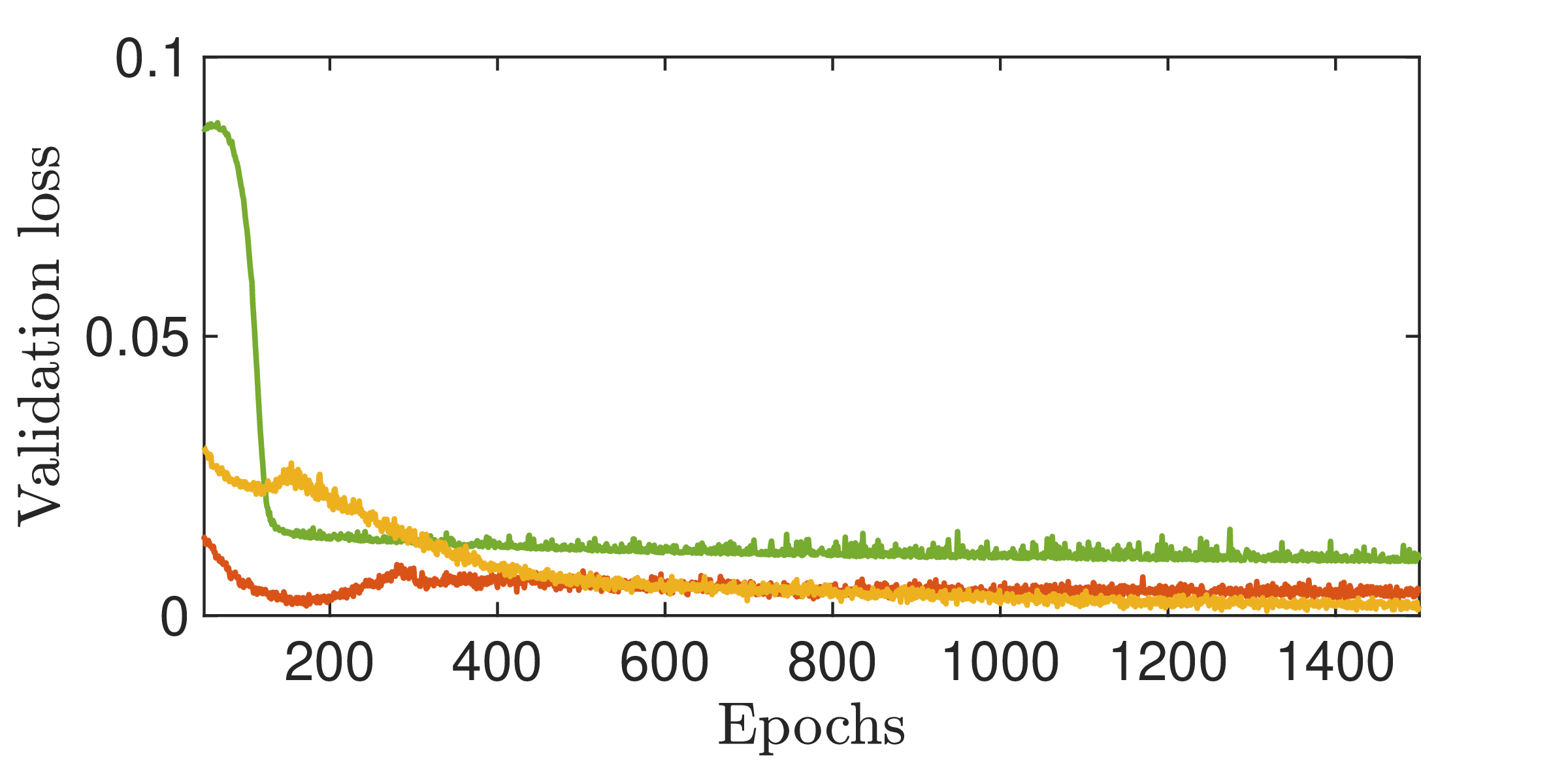}
	\caption{Validation loss comparison: standard NNARX (green line) vs CA-NNARX with $\delta$ISS condition (red line) vs CA-NNARX without $\delta$ISS condition (yellow line).}
	\label{fig:loss_comparsion}
\end{figure}

Figure \ref{fig:loss_comparsion} shows the evolution of the validation metric (i.e., the simulation MSE over the validation dataset, $\mathcal{L}(\mathcal{I}_{\val}, \Phi)$) during the training procedure of three models.
Although the number of training epochs required and the wall-clock time are comparable (around $120$ minutes on an i7 processor with 16GB of RAM), it can be noted that the CA-NNARX model achieves consistently better performances than the black-box NNARX. 
\REV{The FIT index scored by the models on the independent test set is reported in Table~\ref{tab:comparison_fit}.}

\begin{table}[t]
\centering
\caption{Comparison of FIT index of three models}
\label{tab:comparison_fit}
\begin{tabular}{c|ccccc}
\toprule
 \textbf{Models} & $h_1$ & $h_2$ & $h_3$ & $h_4$ & overall average \\ \midrule
$\delta$ISS CA-NNARX &92.1 & 87.1 & 91.1&80.8& $87.8$\\
CA-NNARX &96.2 & 88.6 & 92.2&85.8& $\bm{90.7}$\\
NNARX  &89.8 & 74.1 & 78.1&59.6& $75.4$\\
\bottomrule
\end{tabular}
\end{table}

\REV{Note that the CA-NNARX without $\delta$ISS condition performs slightly better than $\delta$ISS CA-NNARX on the validation dataset, likely due to the conservativeness of \eqref{eq:delta_iss_condition}.
However, as we will show later in this section, when the control architecture is deployed to the real (noisy) apparatus, the stability and robustness conferred to the model by the $\delta$ISS property will be key to achieve satisfactory closed-loop performance.}

\subsection{\REVV{Experimental validation of the proposed IMC law}}\label{sec:close}
Based on the identified $\delta$ISS CA-NNARX model, the IMC scheme proposed in Section \ref{sec:controldesign} has been implemented and tested on the real lab apparatus depicted in Figure \ref{fig:4tank}.

To this end, we first verified that Assumption~\ref{ass:boundedG} holds for the trained $\delta$ISS CA-NNARX model.
To this end, we computed the largest $\epsilon$ satisfying  \eqref{eq:imc:g_ass} by solving the following optimization problem using CasADi
\begin{equation} \label{eq:optsmall}
		\epsilon^* = \min_{x \in \mathcal{X}}  \big\lvert [ g(x) ]_j \big\lvert, \quad \forall j \in \{ 1, ..., n_u \}.
\end{equation}
Solving \eqref{eq:optsmall} for the trained $\delta$ISS CA-NNARX model yields $\epsilon^* = 0.24$, hence Assumption~\ref{ass:boundedG} is fulfilled.

\REV{To synthesize the IMC scheme, we then need to design the time constant of the model reference block $\mathcal{F}_r$ and of the modeling error filter $\mathcal{F}_m$, see Remark~\ref{rem:imcdesign}.
Both time constants have been set to $\tau_r = 12~s$, in order to smooth the effect of sudden variations of the reference signal and to filter out the measurement noise.
The control law \eqref{eq:imc:inverse} has then been deployed to the control board, and closed-loop experiments have been conducted on the lab apparatus.
}

\begin{figure}
    \centering	\includegraphics[width=0.85\columnwidth]{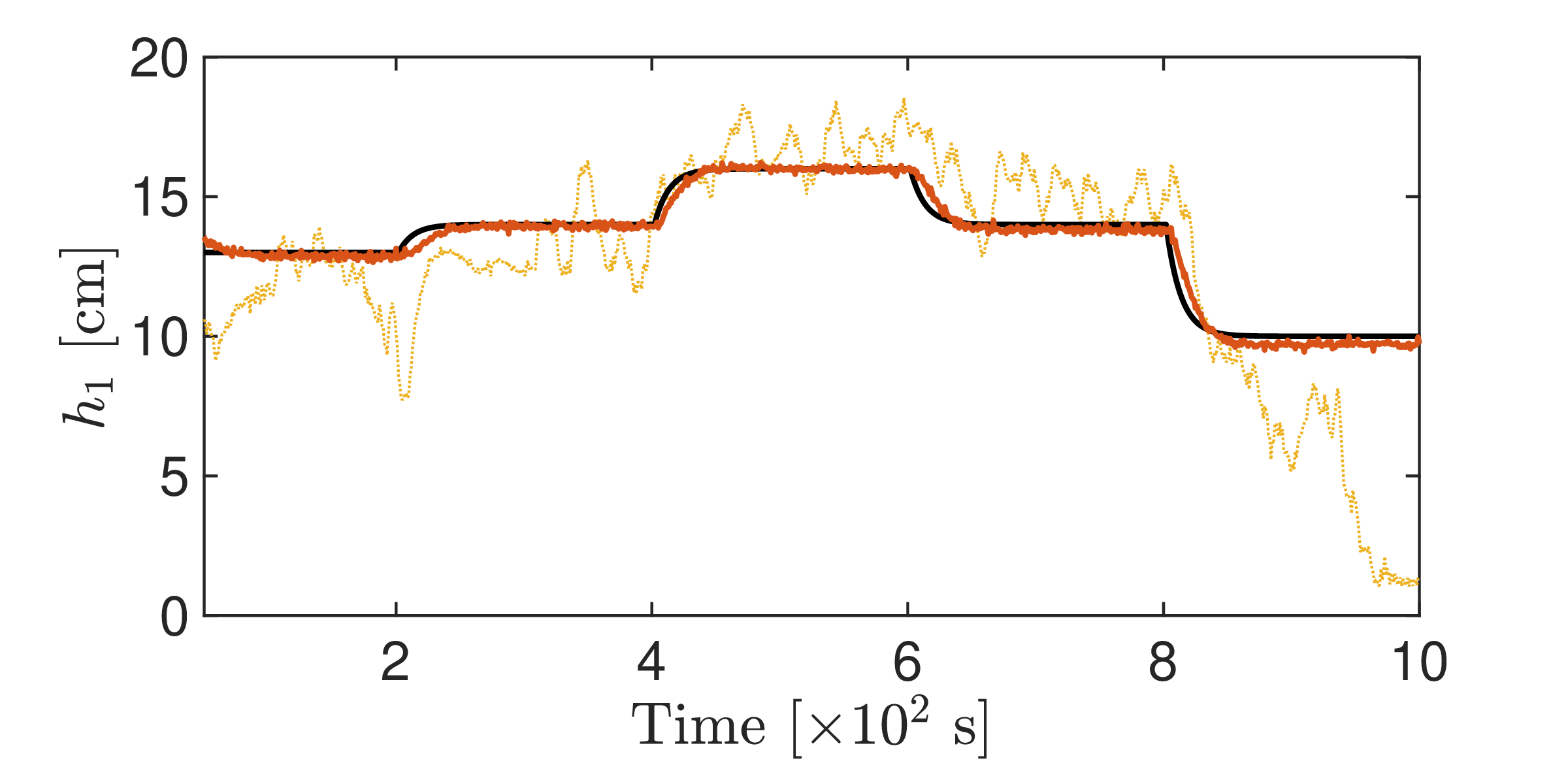}
\includegraphics[width=0.85\columnwidth]{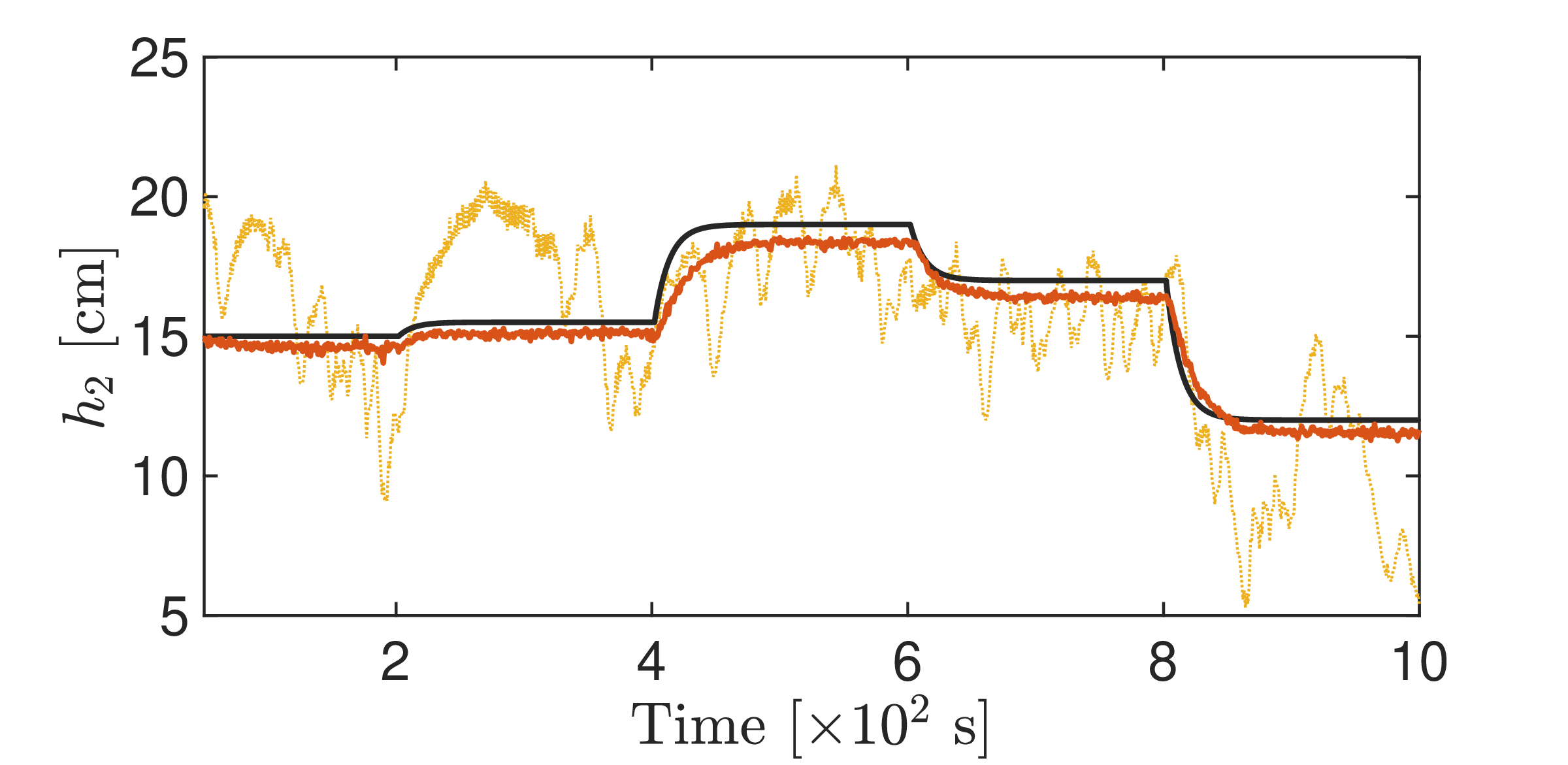}
\includegraphics[width=0.85\columnwidth]{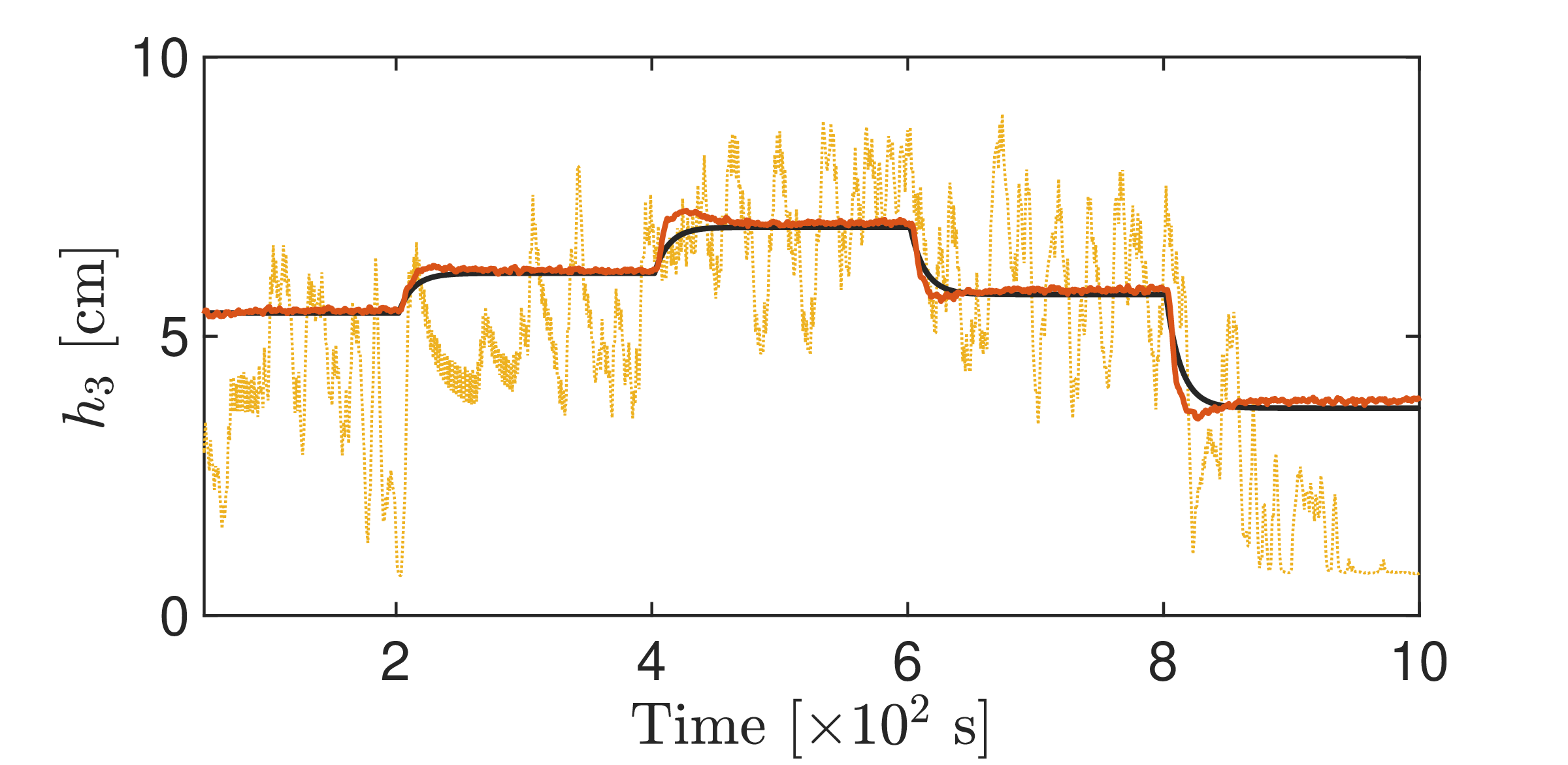}
\includegraphics[width=0.85\columnwidth]{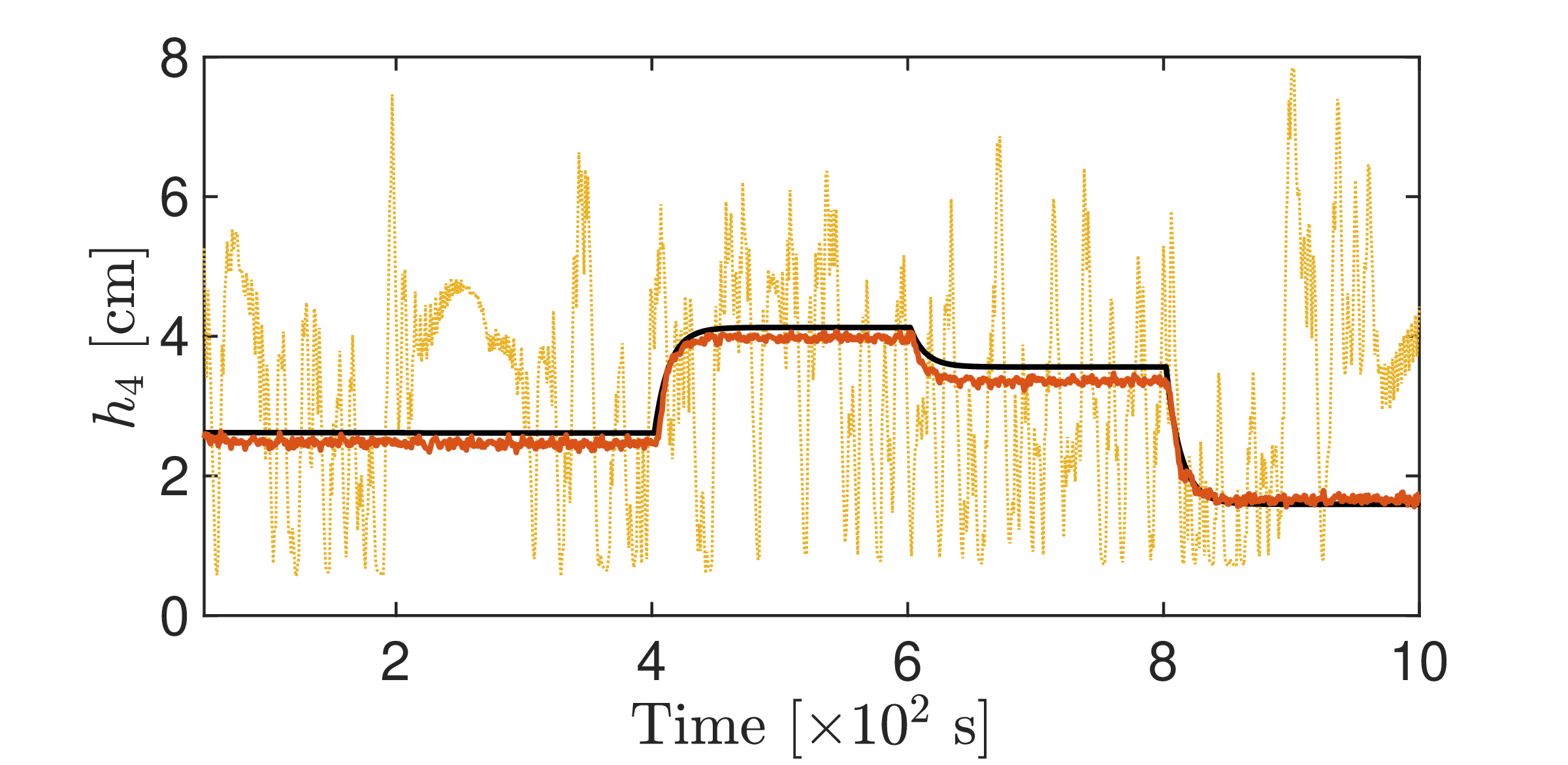}
\caption{\REV{Output tracking performance of the IMC based on the $\delta$ISS (red solid line) and non-$\delta$ISS (yellow dotted line) CA-NNARX models, respectively, compared to the piecewise-constant reference signal filtered by $\mathcal{F}_r$ (black solid line).}}
\label{fig:result_y}
 \end{figure}


\smallskip
\REV{Figure \ref{fig:result_y} shows, for all four outputs, the closed-loop tracking performance achieved by the IMC for random (feasible) piecewise-constant reference signal.
We can first notice that the control scheme ensures the input-output closed-loop stability, see Proposition \ref{prop:iostability}.
In addition, \REVV{ when using the $\delta$ISS CA-NNARX model for IMC design} the closed-loop performance is very satisfactory, even in the presence of measurement noise and of the (inevitable) plant-model mismatch arising when deploying the controller to real systems.}

\REV{The \emph{accuracy} and \emph{limited sensitivity to noise} of this controller based on the $\delta$ISS CA-NNARX model are particularly stunning when compared to those obtained from the same IMC scheme based on the non-$\delta$ISS CA-NNARX model.
This testifies to how learning stable and robust models through the $\delta$ISS property can yield IMC schemes with significantly better dynamic performance, in spite of \REVV{their slightly lower  accuracy compared to} non-$\delta$ISS models, c.f. Table \ref{tab:comparison_fit}}.

\begin{figure}[t]
	\centering
\includegraphics[width=0.85\columnwidth]{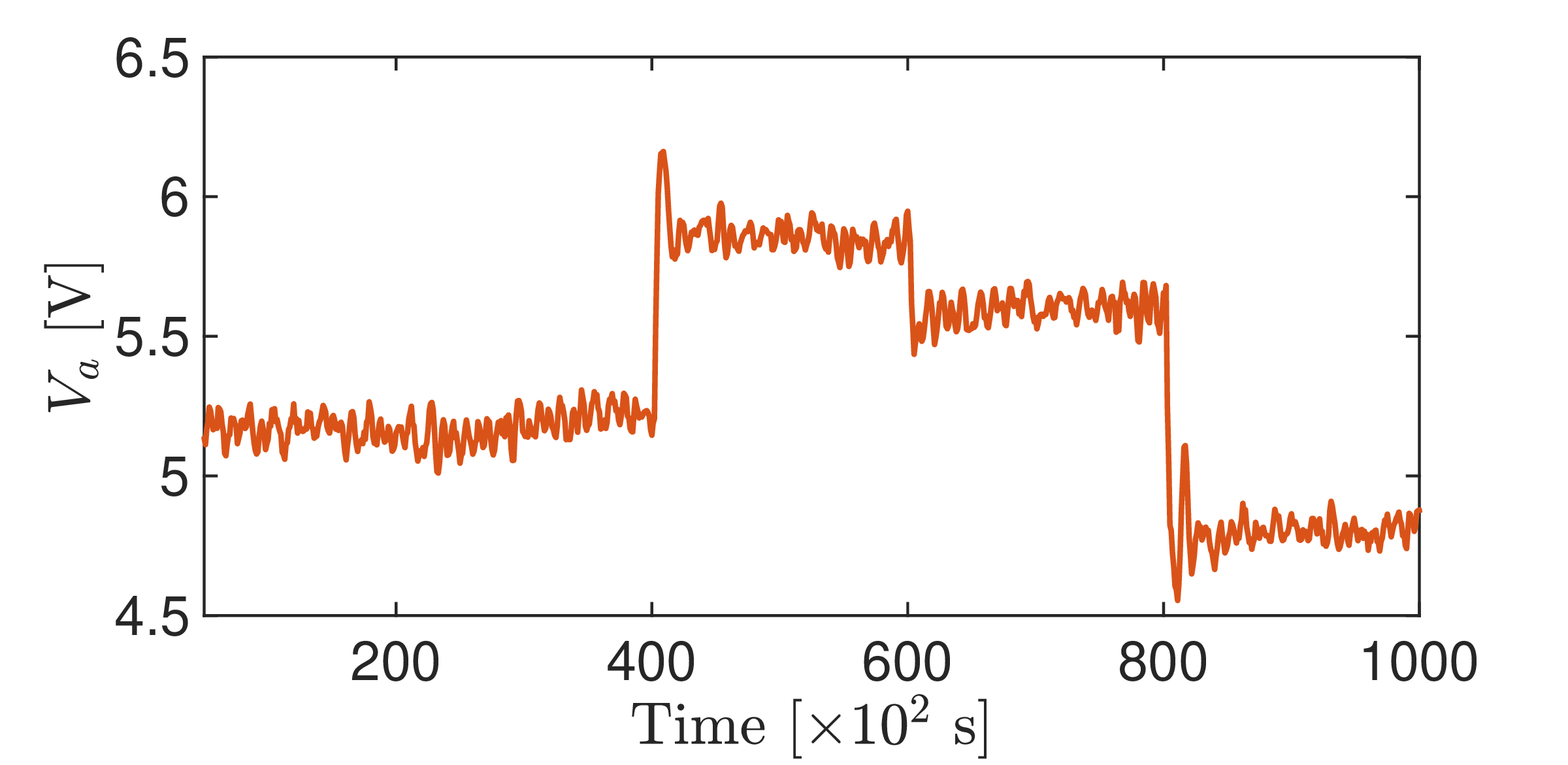}	\includegraphics[width=0.85\columnwidth]{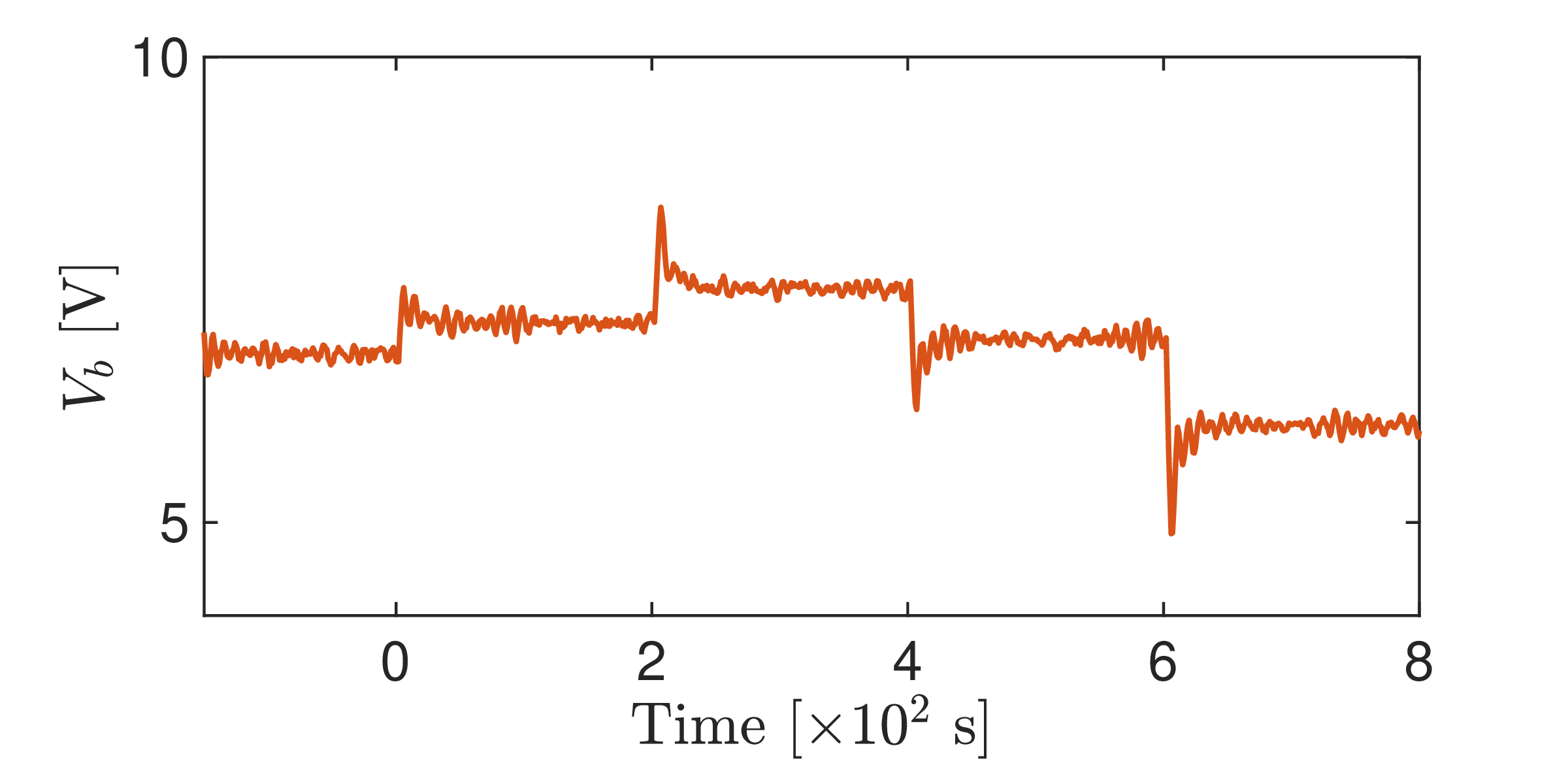}
	\caption{Input $V_a$ (top) and $V_b$ (bottom) applied by IMC scheme \REV{based on the $\delta$ISS CA-NNARX model.}}
    \label{fig:inputresult}
\end{figure}

\smallskip
\REV{The control action applied by the IMC scheme based on the $\delta$ISS CA-NNARX model is depicted in Figure \ref{fig:inputresult}, while
the control action applied by the IMC based on the non-$\delta$ISS CA-NNARX has been omitted for clarity.
Note that the input saturation constraint \eqref{serbatoi_constraints} is always satisfied.}


\subsection{Comparison with MPC in simulation}\label{sec:comparempc}

\REV{After investigating the performance of the IMC scheme based on the $\delta$ISS CA-NNARX  model, we compare the performance, in terms of tracking accuracy and computational burden, of such a scheme with that of an MPC.  
The comparison with this latter strategy is motivated by the fact that MPC constitutes nowadays a standard control strategy, as it allows to embedd nonlinear models as well as constraints on inputs, states, and outputs.}

\REV{Due to software and hardware limitations of the laboratory apparatus, the nonlinear MPC required to handle the nonlinear CA-NNARX model \eqref{eq:model:compact_model} could not be deployed to the control board. 
This comparison has hence been conducted in simulation, meaning that a new $\delta$ISS CA-NNARX model has been identified based on the data collected by simulating the plant  \eqref{serbatoi} with a MPRS signal.
This yielded a model with a FIT index of $93.65\%$.}

\REV{Then, the following controllers have been tested in closed-loop on the same simulator: (\emph{a}) the proposed IMC scheme \eqref{eq:imc:inverse}, with blocks $\mathcal{F}_r$ and $\mathcal{F}_m$ designed as in Section~\ref{sec:close}; (\emph{b}) a nonlinear MPC designed as described in Appendix~\ref{app2}, which serves as a baseline for comparison.}

\begin{figure}[]
    \centering
\includegraphics[width=0.85\columnwidth]{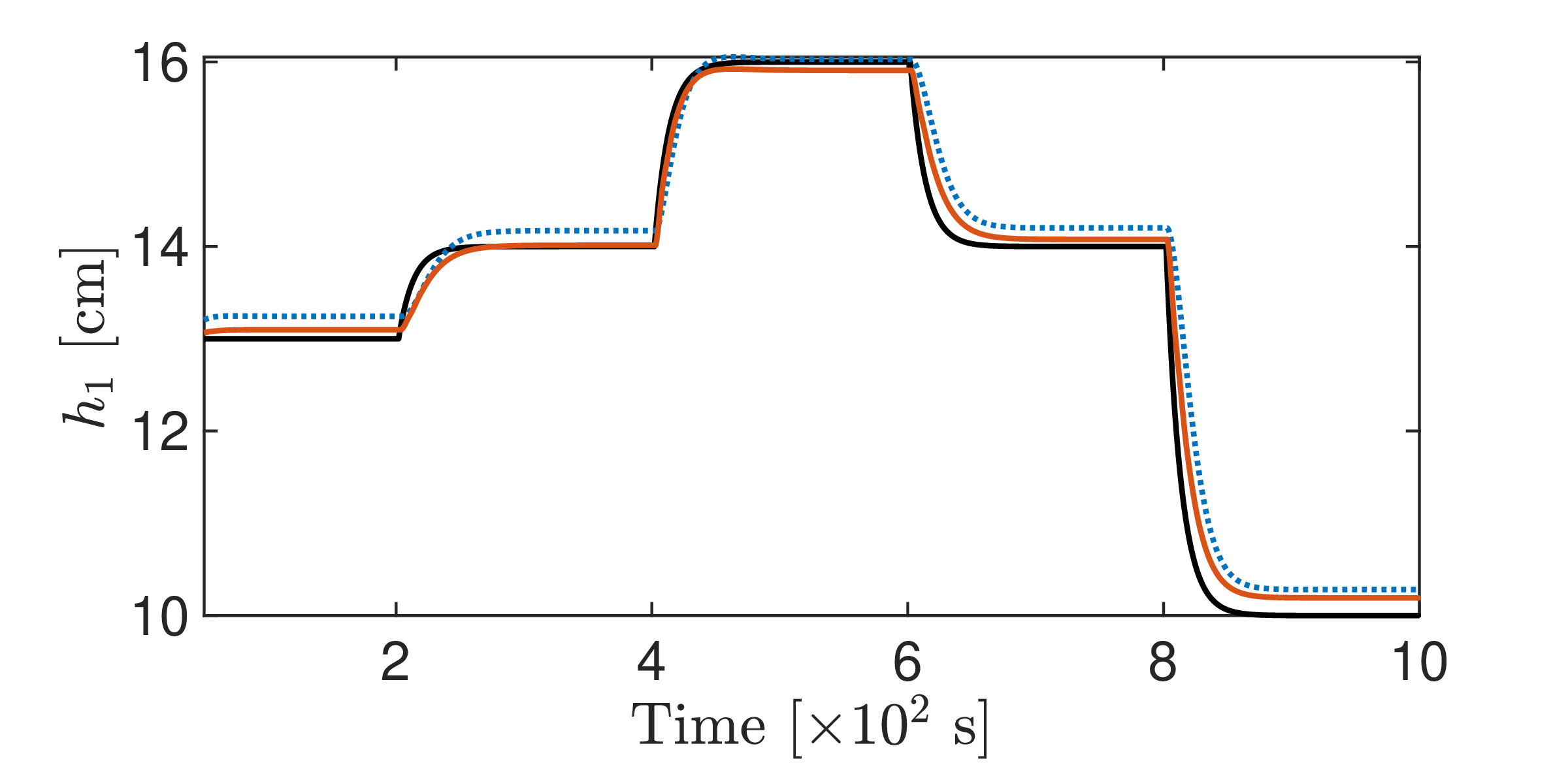}
\includegraphics[width=0.85\columnwidth]{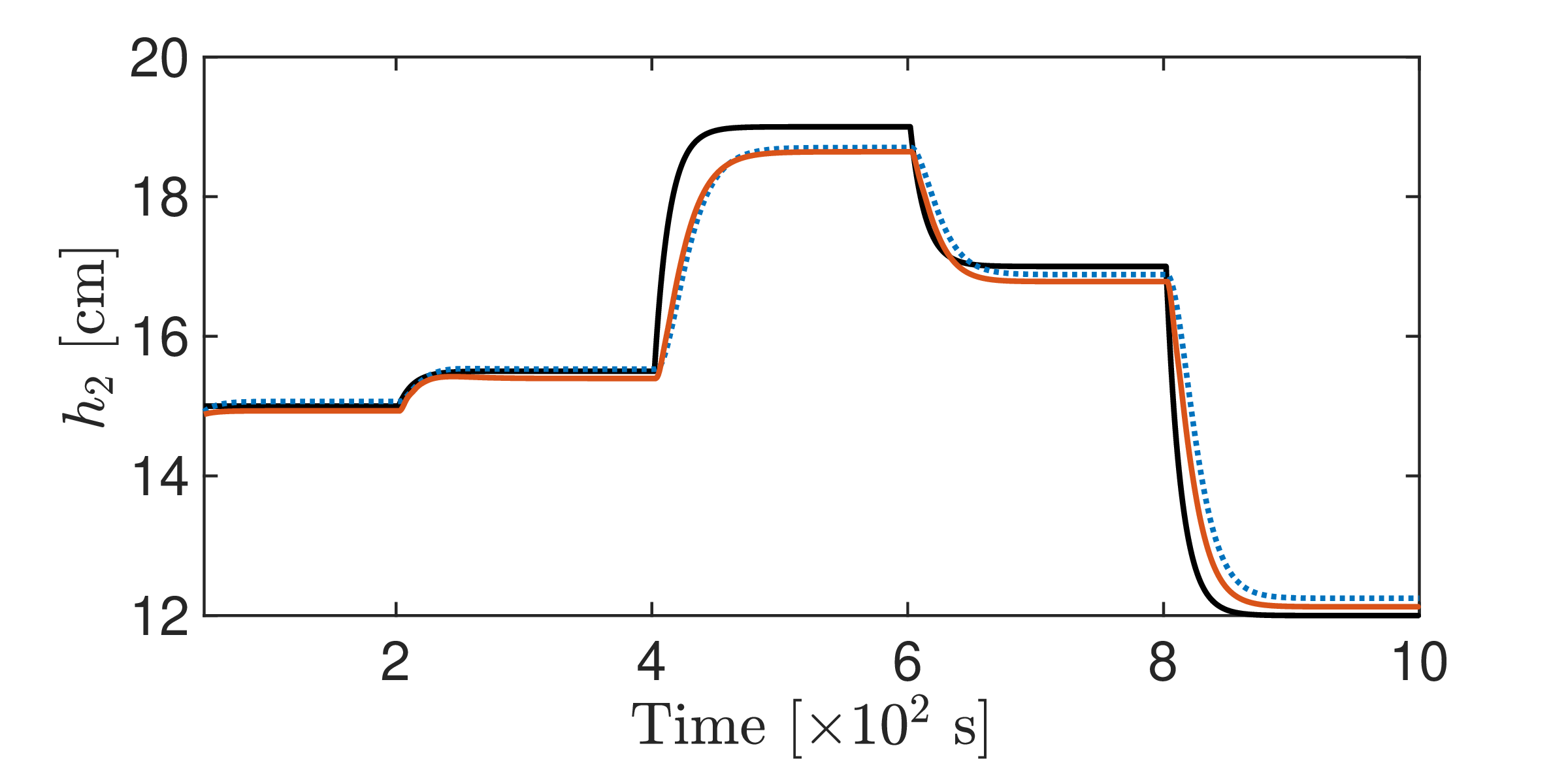}
\includegraphics[width=0.85\columnwidth]{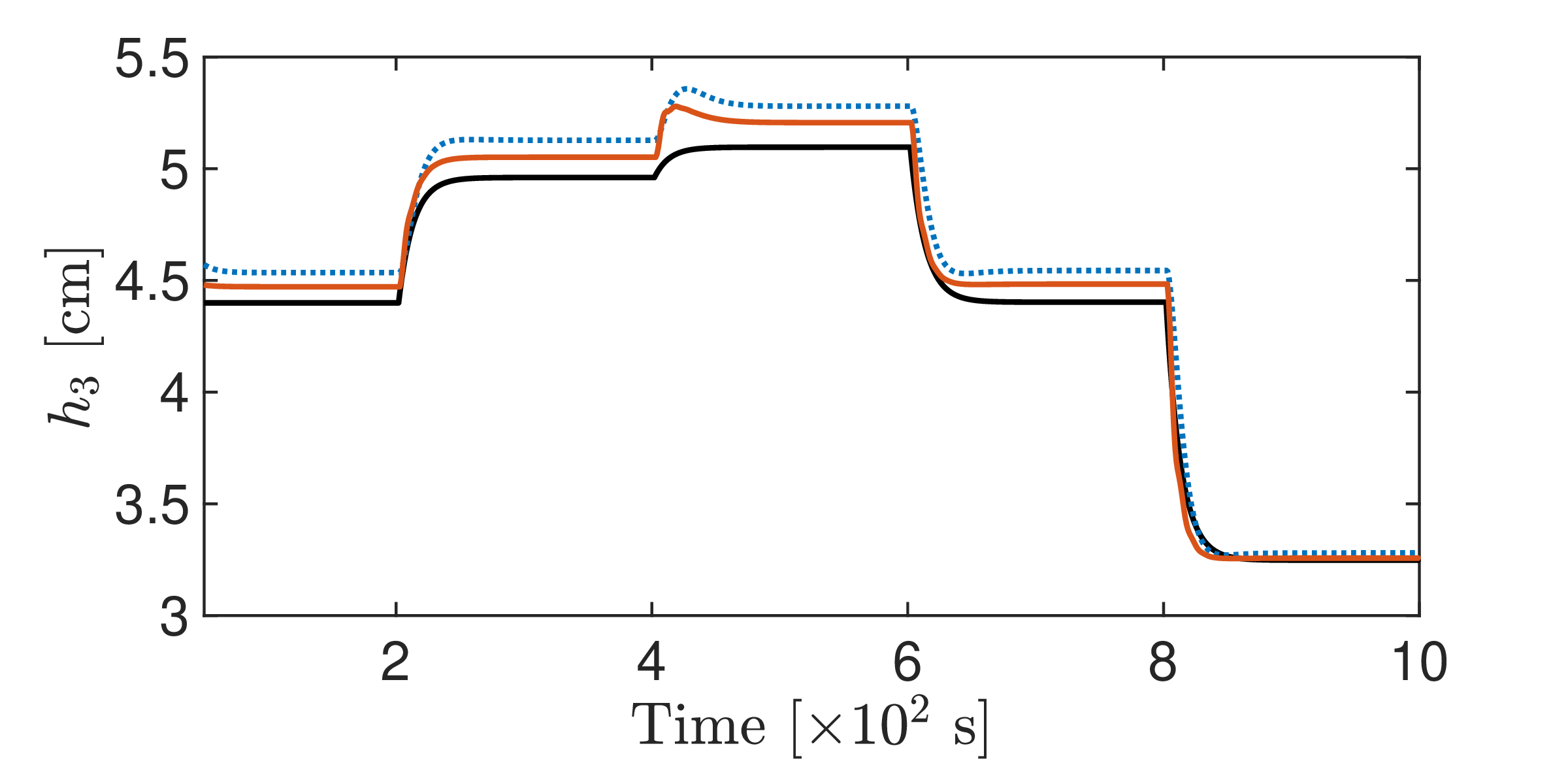}
\includegraphics[width=0.85\columnwidth]{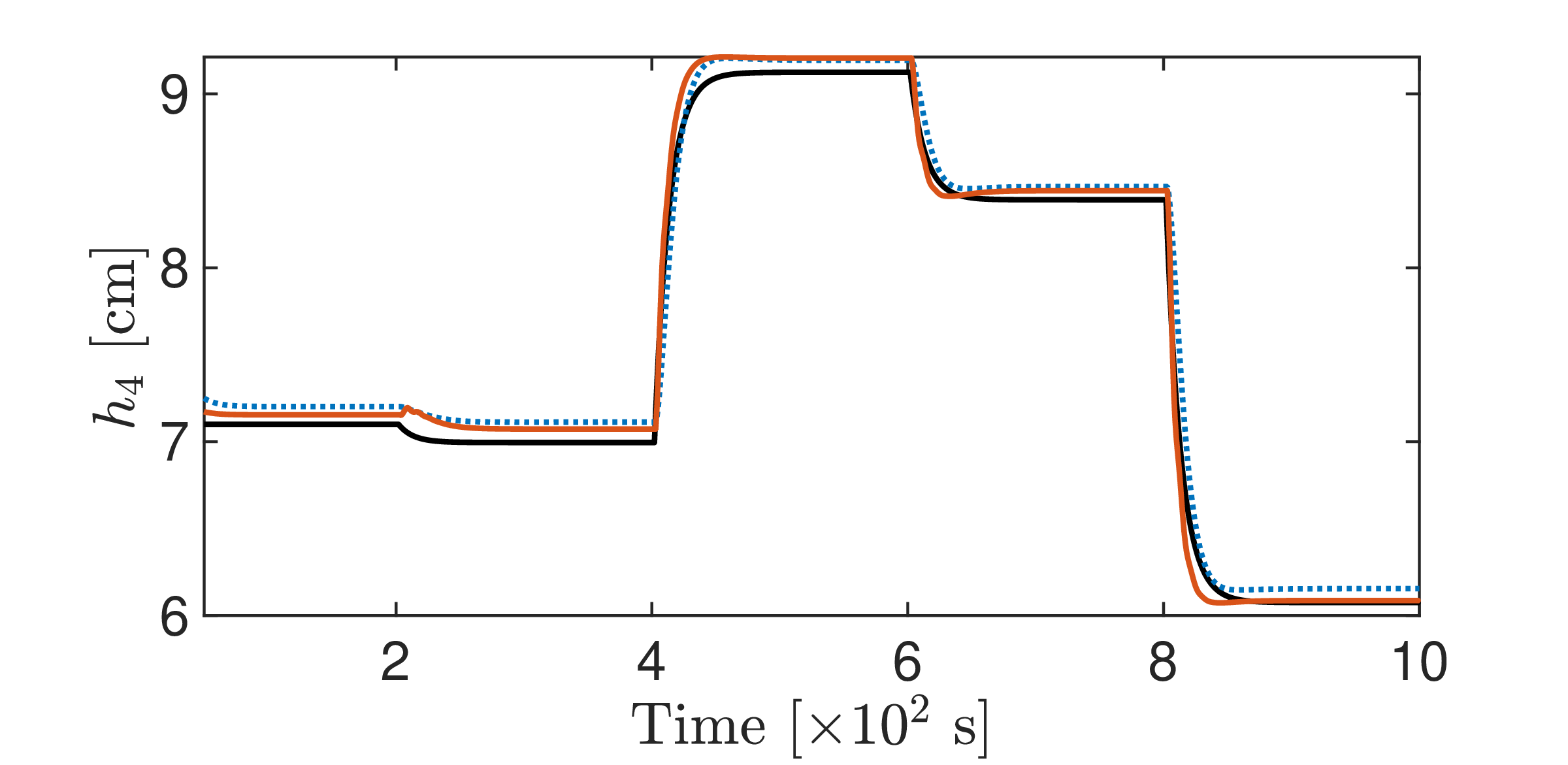}
\caption{Output tracking performances of the proposed $\delta$ISS IMC (red solid line) vs MPC (blue dotted line) vs piecewise-constant reference signal filtered by $\mathcal{F}_r$ (black solid line) for the four output signals.}
\label{fig:simresult_y}
\end{figure}

\smallskip
Figure \ref{fig:simresult_y} shows the closed-loop output tracking performances in simulation achieved by the two controllers.
Both control schemes display closed-loop stability, thanks to the enforced $\delta$ISS.

\begin{table}[b]
\centering
\caption{Comparison of MPC and IMC}
\label{tab:simcomparison}
\begin{tabular}{ccc}
\toprule
 & \textbf{MPC}& \textbf{IMC}  \\ \midrule
RMSE $h_1 [cm]$ & 0.0232 & $\bm{0.0078}$\\
RMSE $h_2 [cm]$ & 0.1306 & $\bm{0.0394}$\\
RMSE $h_3 [cm]$& 0.0145 & $\bm{0.0071}$\\
RMSE $h_4 [cm]$ & 0.2539 & $\bm{0.1506}$\\
Average computational time [s] &3.0645 &$\bm{4.68 \times 10^{-5}}$\\
\bottomrule
\end{tabular}
\end{table}

Note that although the performances achieved by the two schemes are comparable, the root means square of the output tracking error, reported in Table \ref{tab:simcomparison}, reveals slightly better performances of the IMC. 
\REV{This is likely motivated by the fact that the IMC scheme directly exploits model error feedback, whereas MPC requires a model-based state observer, which, in the presence of plant-model mismatch, yields less reliable state estimates.}

\begin{figure}[t]
    \centering
\includegraphics[width=1\columnwidth]{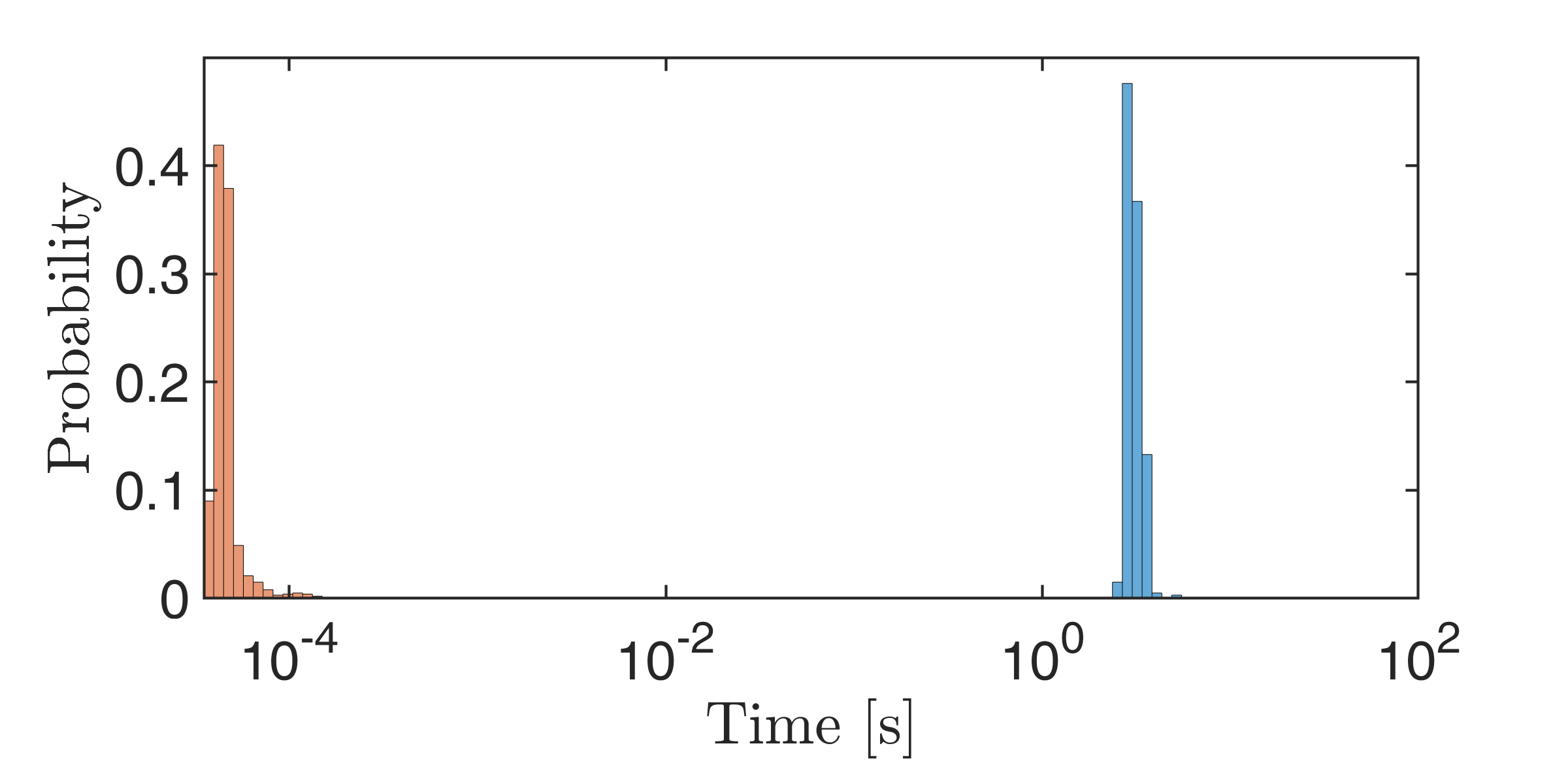}
    \caption{Empirical probability distribution of the computational burden at each control time-step of IMC (red) and of MPC (blue) on the same histogram.}
    \label{fig:simcomputation}
	\vspace{2mm}
\includegraphics[width=0.85\columnwidth]{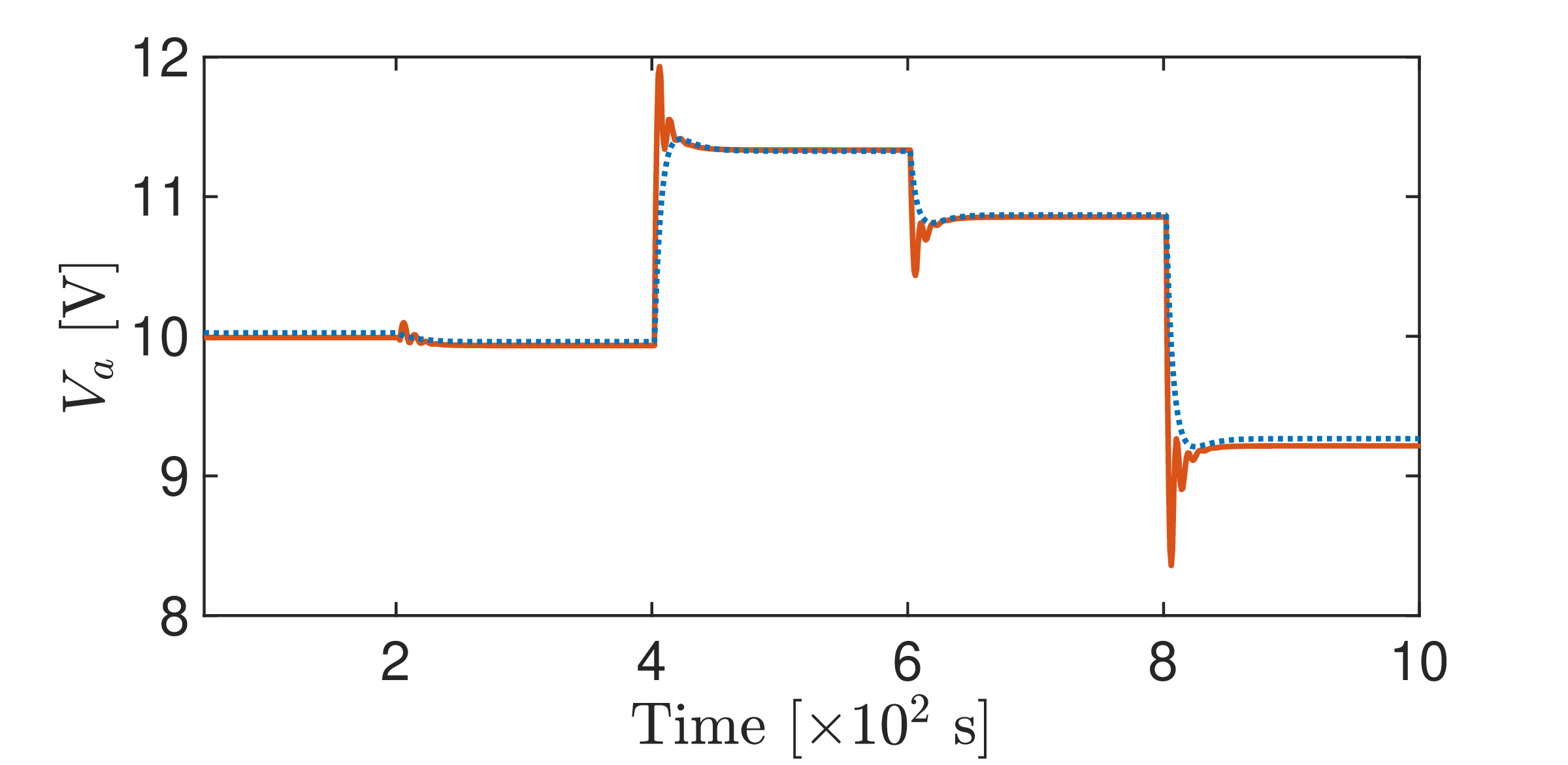}	\includegraphics[width=0.85\columnwidth]{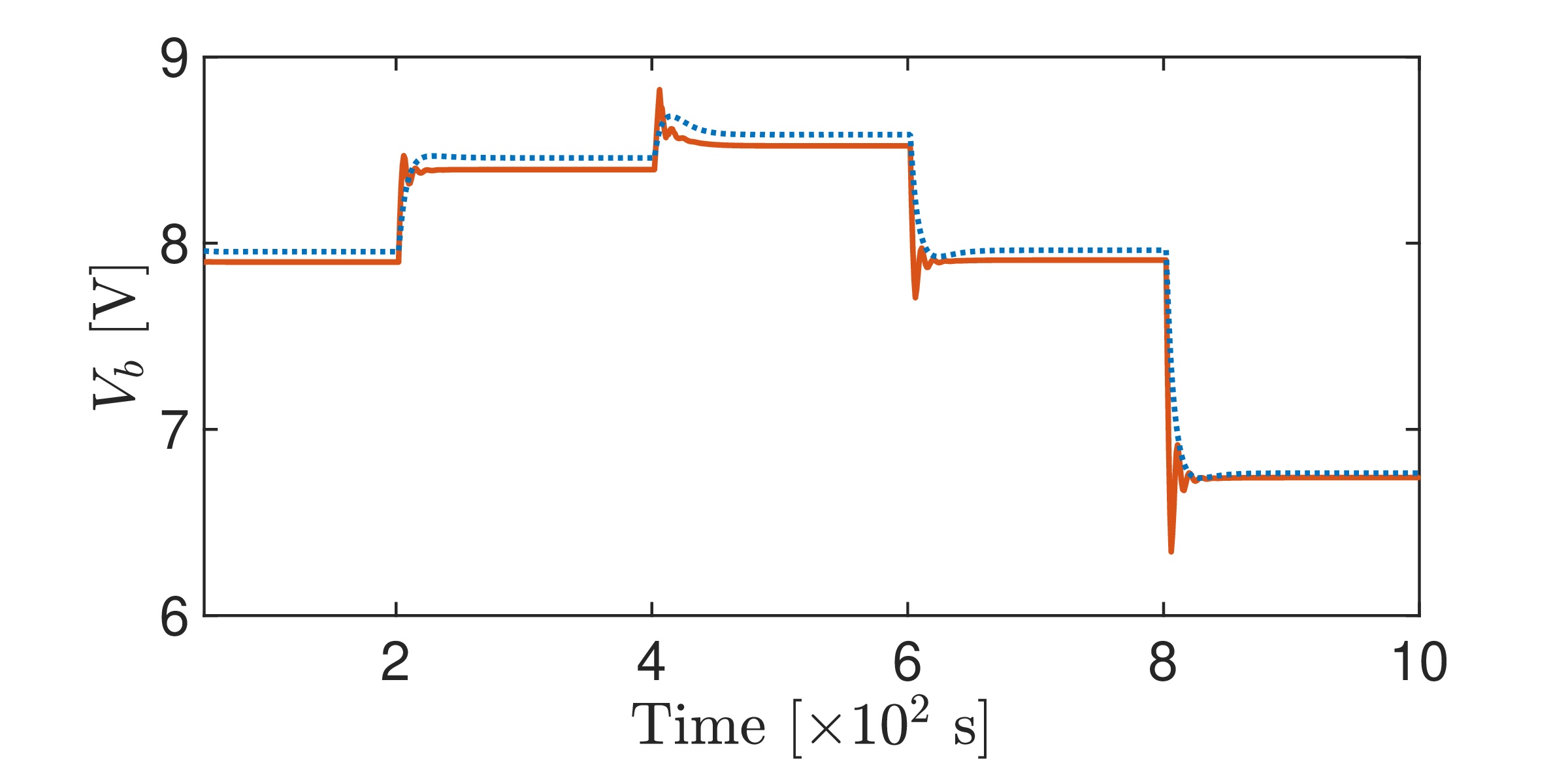}
	\caption{Input $V_a$ (top) and $V_b$ (bottom) of MPC (blue dotted line) vs IMC (red solid line).}
    \label{fig:siminputresult}
\end{figure}

\REV{The great advantage of IMC lies however in its online computational efficiency. Unlike MPC, IMC does not require solving any online optimization problem, but rather only evaluating the explicit control law \eqref{eq:imc:inverse}.
This allows the computational cost of IMC to be five orders of magnitude smaller than that of MPC, see Table~\ref{tab:simcomparison}, and to be easily implemented on the control board of the laboratory apparatus, as described in the previous section.
To support this claim, the empirical distribution of the computational time of the two control laws at each time instant is shown in Figure~\ref{fig:simcomputation}.}

\REV{
Finally, Figure~\ref{fig:siminputresult} shows the control action applied throughout the simulation experiments.
Although MPC demonstrates a better moderation of control during transients, both schemes are able to meet the saturation of the control variable \eqref{serbatoi_constraints}.}

\vfill

\section{Conclusion}\label{sec:conclusion}
In this paper, an Internal Model Control (IMC) scheme for data-driven control affine models has been designed.
We propose a Control-Affine Neural NARX (CA-NNARX) structure that allows to match the structure of the system and admits an explicit inverse, which strongly simplifies the IMC synthesis phase.
The stability properties of CA-NNARX models have been investigated, and a strategy for learning provenly-stable CA-NNARX models has been proposed.
The model's stability allows, under mild assumptions, to derive closed-loop input-output stability guarantee for the proposed IMC scheme.

The proposed approach has been tested on a real Quadruple Tank system and compared to a model predictive controller, showing a significantly smaller computational burden and similar output tracking performances.
Future research directions will include the robustness analysis of the proposed approach, the development new algorithms for the estimation of bounds on the modeling error, and robust control algorithms for CA models.
\REVV{In addition, research efforts will be devoted to retrieve a sufficient condition for $\delta$ISS that scales better with respect to the regression horizon $H$ and that is less conservative. }

\begin{appendices}
\section{Proof of Theorem \ref{thmdISS}}\label{app1}
Before proving Theorem \ref{thmdISS}, let us introduce the following instrumental notions.

\begin{definition} [$\delta$ISS-Lyapunov function \cite{bayer2013discrete}]
A continuous function $V: \mathbb{R}^n \rightarrow \mathbb{R}_{\geq 0}$ is said to be a $\delta$ISS-Lyapunov function for system \eqref{eq:model:compact_model} if there exist functions $\psi_4, \psi_2, \psi_3, \psi_4 \in \mathcal{K}_{\infty}$ such that, for any $x_{a,k} \in \mathcal{X}$ and $ x_{b,k}  \in \mathcal{X}$, and any $u_{a,k} \in \mathcal{U}$ and $u_{b,k} \in \mathcal{U}$, it holds that
\begin{equation}\label{eq:deltaiss:lyapunov_function}
\begin{aligned}
\psi_1(\left \| x_{a,k}  - x_{b,k}\right \|) \leq V(x_{a,k}, x_{b,k}) \leq& \psi_2(\left \| x_{a,k}- x_{b,k} \right \|), \\
V(x_{a, k+1, x_{b, k+1}}) - V(x_{b, k}, x_{b, k}) \leq& -\psi_3(\left \| x_{a,k}-x_{b,k}\right \|) \\
&\quad + \psi_4(\left \| u_{a,k}-u_{b,k}\right \|)
\end{aligned}
\end{equation}
with $x_{a,k+1} = F(x_{a,k}, u_{a,k}; \bm{\Phi})$, $x_{b,k+1} = F(x_{b,k}, u_{b,k}; \bm{\Phi})$.
\end{definition}
Then, the following relationship exists between the system's $\delta$ISS property and the existence of  $\delta$ISS-Lyapunov functions.
\begin{lemma} [$\delta$ISS \cite{bayer2013discrete}]  \label{lemma:deltaiss}
If system \eqref{eq:model:compact_model} admits a continuous $\delta$ISS-Lyapunov function, then it is $\delta$ISS.
\end{lemma}

Lemma \eqref{lemma:deltaiss} thus allows us to assess $\delta$ISS property of the system by finding a suitable $\delta$ISS-Lyapunov function.
To this end, let us start by noticing that, taking $Q = I$, the matrix $P = \text{diag}(I, 2 \cdot I, 3 \cdot I,..., N \cdot I)$ solves the discrete-time Lyapunov equation $A'PA-P = -Q$, where $A$ is defined in \eqref{eq:model:matrices}.
We thus consider the $\delta$ISS-Lyapunov function candidate $V(x_a,x_b) = (x_a-x_b)'P(x_a-x_b)$.
Note such Lyapunov function candidate satisfies
\begin{equation}
    \lambda_{\text{min}}(P)\left \|x_a-x_b \right \|_2^2 \leq V(x_a,x_b) \leq \lambda_{\text{max}}(P)\left \| x_a - x_b \right \|_2^2
\end{equation}
where $\lambda_{\text{min}}(P) = 1$ and $\lambda_{\text{max}}(P) = H$ denote the minimum and maximum singular value of matrix $P$, respectively.
Then, it follows that
\begin{equation}\label{eq:delta_ISS}
\scalemath{0.75}{
\begin{aligned}
    & V(x_{a, k+1}, x_{b, k+1}) - V(x_{a, k},x_{b, k}) \\
    & = \big[ F(x_{a, k}, u_{a, k}) - F(x_{b, k}, u_{b, k}) \big]^\prime \cdot P \cdot \big[ F(x_{a, k}, u_{a, k}) - F(x_{b, k}, u_{b, k}) \big] \\
    &\qquad  - \big( x_{a, k} - x_{b, k} \big)^\prime \cdot P \cdot \big( x_{a, k} - x_{b, k} \big) \\
    &  = \big[ A x_{a, k} +B_u u_{a, k} + B_x \eta(x_{a, k}, u_{a, k})  - A x_{b,k} - B_u u_{b, k} - B_x \eta(x_{b, k}, u_{b, k}) \big]^\prime\\
    & \quad \cdot P \cdot \big[ A x_{a, k} +B_u u_{a, k} + B_x \eta(x_{a, k}, u_{a, k})  - A x_{b,k} - B_u u_{b, k} - B_x \eta(x_{b, k}, u_{b, k}) \big] \\
   &\quad - (x_{a, k} - x_{b, k})^\prime \cdot P \cdot (x_{a, k} - x_{b,k})
\end{aligned}}
\end{equation}
where the dependency of $F(x, u)$ and $\eta(x, u)$ upon the weights $\Phi$ is omitted for compactness.
Let now $\Delta x_k = x_{a,k} - x_{b, k}$, $\Delta u_k = u_{a, k} - u_{b, k}$, $\Delta \eta_k = \eta(x_{a, k}, u_{a, k}) - \eta(x_{b, k}, u_{b, k})$, and $\Delta V = V(x_{a,k+1}, x_{b,k+1})- V(x_{a,k}, x_{b,k})$.
Then, recalling \eqref{eq:model:statespace}, equation \eqref{eq:delta_ISS} can be reformulated as follows
\begin{equation}\label{eq:delta_ISS_2}
\begin{aligned}
      \Delta V  =& \Delta x_k^\prime (A'PA - P) \Delta x_k + \Delta u_k^\prime B_u^\prime  P B_u \Delta u_k \\ &
      + 2 \Delta x_k^\prime A^\prime P B_u \Delta u_k  +\Delta \eta_k^\prime B_x^\prime P B_x \Delta \eta_k  \\
      &+ 2 \Delta x^\prime A' P B_x \Delta \eta_k + 2\Delta u_k^\prime B_u^\prime  P B_x \Delta \eta_k
\end{aligned}
\end{equation}
Recalling that $Q=I$, in view of the structure of the matrices $A$, $B_u$, $B_x$, and $P$, one can easily show that
\begin{equation} \label{eq:proof:matrix_products}
  \begin{split}
    A'PA-P &= -I,\\
    B_u'PB_x &= 0, \\
    B_uPB_u &= H,
  \end{split}
\quad\quad
  \begin{split}
    A'PB_u &= 0,\\
    A'PB_x &= 0,\\
    B_xPB_x &= H.
  \end{split}
\end{equation}

In view of \eqref{eq:proof:matrix_products}, \eqref{eq:delta_ISS_2} reads as
\begin{equation}\label{eq:delta_iss_final}
	\Delta V= -\norm{\Delta x_k}^2_2 + H \norm{\Delta u_k}^2_2 + H\norm{\Delta \eta_k}^2_2
\end{equation}
To show the existence of functions $\psi_3(\| \Delta x_k \|_2)$ and $\psi_4(\| \Delta u_k \|_2)$ such that \eqref{eq:deltaiss:lyapunov_function} is fulfilled, let us bound the term $\left \| \Delta \eta_k \right \|_2^2$.
Recalling \eqref{eq:model:controlaffine_nn}, it holds that

\begin{equation} \label{eq:delta_iss_delta_eta}
\scalemath{0.75}{
    \begin{aligned}
    \| \Delta \eta_k \|^2_2 &= \| \eta(x_{a,k}, u_{a, k})-\eta(x_{b, k}, u_{b, k}) \|^2_2  \\
    & = \big\| W_0 f(x_{a, k}) + U_0 g(x_{a, k}) \otimes u_{a, k} - W_0 f(x_{b, k})- U_0 g(x_{b, k}) \otimes u_{b, k} \big\|^2_2 \\
    & \leq \Big[ \| W_0 \|_2 \, \big\| f(x_{a, k})-f(x_{b, k}) \big\|_2 \\
    & \qquad +  \| U_0 \|_2 \, \big\| g(x_{a, k}) \otimes u_{a, k} - g(x_{b, k}) \otimes u_{b, k} \pm g(x_{a, k}) \otimes u_{b, k} \big\|_2 \Big]^2
    \end{aligned}}
\end{equation}
where $g(x_a) \otimes u_b $ has been summed and subtracted.
Let us point out, at this stage, that both $f(\cdot)$ and $g(\cdot)$ are Lipschitz-continuous, because such networks are defined as sequences of affine transformations followed by element-wise Lipschtiz-continuous activation functions, see \eqref{eq:model:ffnn_f} and \eqref{eq:model:ffnn_g}.
Therefore, it follows that
\begin{subequations} \label{eq:delta_iss_f}
	\begin{align}
			\| f(x_{a,k}) -f(x_{b,k}) \|_2  &\leq \bigg( \prod_{i=1}^{L}\Lambda_{i}  \| W_i \|_2 \bigg)  \| \Delta x_k \|_2 \\
	\| g(x_{a,k}) -g(x_{b,k}) \|_2  &\leq \bigg( \prod_{j=1}^{M}\tilde{\Lambda}_{j}  \| U_i \|_2 \bigg)  \| \Delta x_k \|_2
	\end{align}
\end{subequations}

Recalling that, given vectors $v$ and $z$, it holds that $v \otimes z = v \cdot \text{diag}(z)$, and hence $\| v \otimes z \|_2 \leq \| v \|_2 \| \text{diag}(z) \|_2 = \| v \|_2 \| z \|_\infty$.
In light of this, the last term of \eqref{eq:delta_iss_delta_eta} can be bounded as
\begin{equation} \label{eq:delta_iss_last_term}
\scalemath{0.75}{
    \begin{aligned}
    \big\| g(x_{a,k})&\otimes u_{a,k} -g(x_{b, k})\otimes u_{b, k} \pm g(x_{a, k}) \otimes u_{b, k} \big\|_2 \\
    &\leq \big\| \big( g(x_{a, k}) - g(x_{b, k}) \big) \otimes u_{b, k} \big\|_2  + \| g(x_{a, k}) \otimes (u_{a, k} - u_{b, k}) \|_2\\
    &\leq \left ( \prod_{j=1}^{M} \tilde{\Lambda}_j \norm{U_j}_2\right ) \| \Delta x_k \|_2 \cdot \sup_{u_b \in \mathcal{U}} \| u_b \|_\infty + \sup_{x_a \in \mathcal{X}} \norm{g(x_a)}_{\infty} \cdot \| \Delta u_k \|_2
    \end{aligned}}
\end{equation}
In light of Assumption \ref{ass:input}, $\| u_b \|_\infty = 1$. Moreover, because of the radial boundedness of $\varsigma_j(\cdot)$, it holds that $\sup_{x_a} \norm{g(x_a)}_{\infty} =1$.
Then, the upper bound \eqref{eq:delta_iss_last_term} reads as
\begin{equation} \label{eq:delta_iss_g_final}
\begin{aligned}
   \big\| g(x_{a,k})\otimes& u_{a,k} - g(x_{b, k}) \, \otimes \, u_{b, k} \pm g(x_{a, k}) \otimes u_{b, k} \big\|_2 \\
   &  \leq \left ( \prod_{j=1}^{M}\tilde{\Lambda}_j \norm{U_j}_2 \right ) \| \Delta x_k \|_2+ \| \Delta u_k \|_2
  \end{aligned}
\end{equation}
Thus, in light of \eqref{eq:delta_iss_f} and \eqref{eq:delta_iss_g_final}, by applying the Young inequality on \eqref{eq:delta_iss_delta_eta} we get
\begin{equation} \label{eq:delta_iss_delta_eta_final}
\scalemath{0.75}{
\begin{aligned}
\| \Delta \eta_k \|_2^2 & \leq \Big[ \Big( \| W_0 \|_2 \prod_{i=1}^{L} \Lambda_i \| W_i \|_2 + \| U_0 \|_2 \prod_{j=1}^{M} \tilde{\Lambda}_j \| U_j \|_2 \Big) \| \Delta x_k \|_2 + \| U_0 \|_2  \| \Delta u_k \|_2  \Big]^2 \\
& \leq \left( 1 + \frac{1}{q^2} \right) \Big( \| W_0 \|_2 \prod_{i=1}^{L} \Lambda_i \| W_l \|_2 + \| U_0 \|_2 \prod_{j=1}^{M} \tilde{\Lambda}_j \| U_j \|_2  \Big)^2  \| \Delta x_k \|_2^2 \\
& \qquad + \left(1+ q^2 \right) \| U_0 \|_2^2 \| \Delta u_k \|_2^2
\end{aligned}}
\end{equation}
for any $q \neq 0 $. Applying \eqref{eq:delta_iss_delta_eta_final} to  \eqref{eq:delta_iss_final} we get
\begin{equation}
\scalemath{0.75}{
\begin{aligned}
    \Delta V \leq &- \underbrace{\bigg[ 1 - H \left( 1 + \frac{1}{q^2} \right) \Big( \| W_0 \|_2 \prod_{i=1}^{L} \Lambda_i \| W_l \|_2 + \| U_0 \|_2 \prod_{j=1}^{M} \tilde{\Lambda}_j \| U_j \|_2  \Big)^2 \bigg]}_{\alpha_x} \| \Delta x_k \|_2^2 \\
    &+ \underbrace{H \big[ 1+ \| U_0 \|_2^2 \big( 1 + q^2 \big) \big]}_{\alpha_u} \| \Delta u_k \|_2^2
    \end{aligned}}
\end{equation}

If condition \eqref{eq:delta_iss_condition} is fulfilled, there exists a sufficiently large value of $q$ such that $\alpha_x > 0$, which implies that $V$ is a $\delta$ISS-Laypunov function, with $\psi_3(\| \Delta x_k \|_2) = \alpha_x \| \Delta x_k \|_2^2$ and $\psi_4(\| \Delta u_k \|_2) = \alpha_u \| \Delta u_k \|_2^2$.
Lemma \ref{lemma:deltaiss} hence entails that system \eqref{eq:model:compact_model} is $\delta$ISS.
\qed

\section{\REV{MPC design}}\label{app2}
The standard Nonlinear Model Predictive Controller scheme  adopted as a baseline for benchmarking the proposed IMC scheme is now described.

To this end, let us assume that, for a given reference value $y^o$, there exist $x^o \in \mathcal{X}$ and $u^o \in \mathcal{U}$ such that the triplet $(x^o, u^o, y^o)$ is an equilibrium of system \eqref{eq:model:compact_model}, i.e., $x^o = F(x^o, u^o)$ and $y^o = G(x^o)$.

MPC works by solving, at every time-step $k$, a Finite Horizon Optimal Control Problem (FHOCP), which computes a control sequence that minimizes a cost function that encodes the desired closed-loop performances.
In particular, letting $N_p$ denote the finite prediction horizon over which the predicted closed-loop performances are optimized, and denoted by $\big\{ u_{0 \lvert k}, ..., u_{N_p - 1 \lvert k} \big\}$ the sequence of control actions throughout such horizon, the FHOCP can be stated as follows
\begin{subequations} \label{eq:nomMPC}
	\begin{align}
		\min_{u_{0|k}, ..., u_{N_p-1|k}} &  \sum_{i=0}^{N_p-1} \Big[ \big\| u_{i|k} - u^o \big\|_R^2 + \big\| \REV{\hat{y}_{i|k}} - y^o \big\|_Q^2 \Big]  \label{eq:nomMPC:cost} \\
		\text{s.t.} \quad & \forall i \in \{ 0, ..., N_p-1 \} \nonumber \\
		&  x_{0|k} =x_{k}  \label{eq:nomMPC:x0} \\
		& x_{i+1|k} = F(x_{i|k}, u_{i|k}) \label{eq:nomMPC:dynamics} \\
		& \REV{\hat{y}_{i|k}} = Cx_{i|k}
        \label{eq:nomMPC:output} \\
		& x_{i|k} \in \mathcal{X} \label{eq:nomMPC:state} \\
		& u_{i|k} \in \mathcal{U} \label{eq:nomMPC:actuator} \\
		&  x_{N_p|k} = x^o\label{eq:nomMPC:terminal}
	\end{align}
\end{subequations}
	Note that the CA-NNARX model is embedded as a predictive model via constraints \eqref{eq:nomMPC:dynamics} and \eqref{eq:nomMPC:output}, where the initial condition is fixed to the current measured state via \eqref{eq:nomMPC:x0}.

Constraint \eqref{eq:nomMPC:state} and \eqref{eq:nomMPC:actuator} allow to enforce the satisfaction, throughout the prediction horizon, of state and input constraints, respectively.
A zero-terminal constraint is included, see \eqref{eq:nomMPC:terminal}, to guarantee the nominal closed-loop stability property \cite{rawlings2017model}.

The adopted cost functions penalizes the deviations of input and output trajectories from the equilibrium, see \eqref{eq:nomMPC:cost}, where $R = \text{diag}(0.1,0.1)$ and $Q = \text{diag}(5,5,5,5)$ are positive definite weight matrices.
In the implemented controller, the prediction horizon has been set to $N_p = 10$.

According to the Receding Horizon principle, at time instant $k$, the optimization problem \eqref{eq:nomMPC} is solved, and an optimal control sequence $u_{0|k}^\star, ..., u_{N_p-1|k}^\star$ is retrieved. Then, only the first element $u_{0|k}$ is applied to the plant, i.e., $u_k = u_{0|k}^\star$.
Then, at the successive time instant, the procedure is repeated based on the new measurement state, $x_{k+1}$.
\end{appendices}

\bibliographystyle{IEEEtran}
\bibliography{reference}
\begin{IEEEbiography}[{\includegraphics[width=1in,height=1.25in,clip]{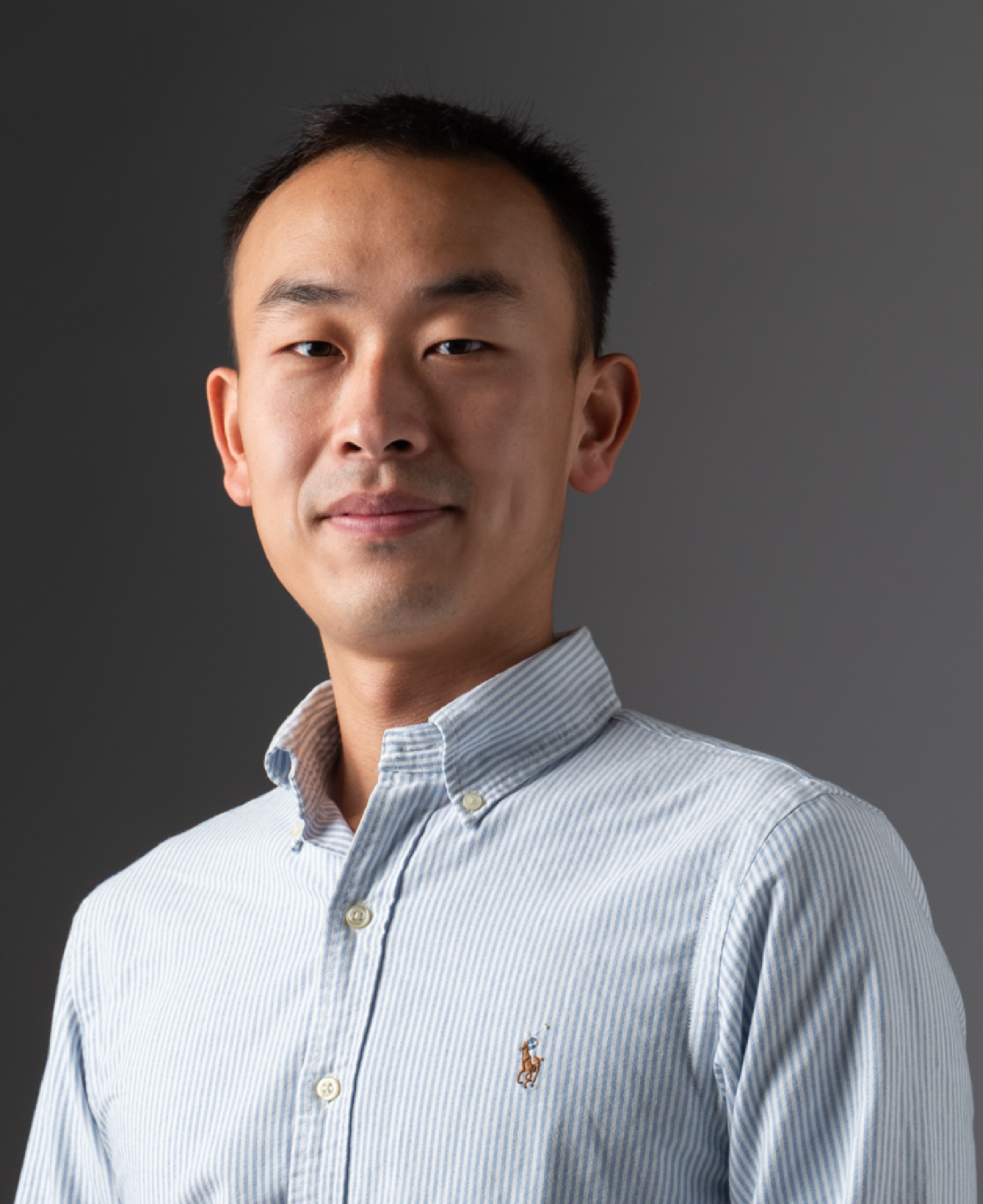}}]{Jing Xie}
graduated with a bachelor’s degree in automation at Southwest Jiaotong University, China, and a master's degree in electrical engineering with a focus on automation and robotics at Technical University of Munich. He did an internship at Robert Bosch on in-car monitoring system using machine learning methods in Hildesheim. Since May 2021, he is a PhD student at Politecnico di Milano under the supervision of Prof. Riccardo Scattolini. He is also an early-stage researcher within the European ELO-X program, which focuses on embedded learning and optimization for the next generation of smart industrial control systems. His main research interests are Physics-based neural networks, data-driven system identification, and Model Predictive Control.
\end{IEEEbiography}
\begin{IEEEbiography}[{\includegraphics[width=1in,height=1.25in,clip,keepaspectratio]{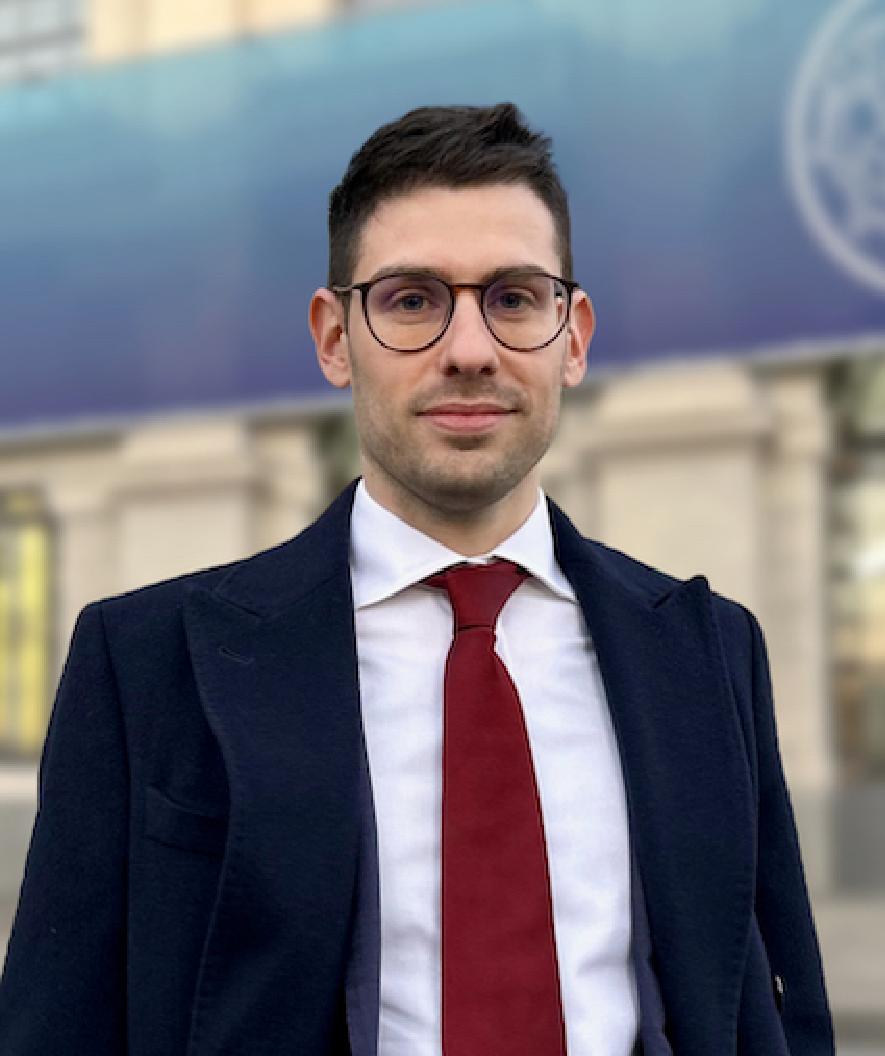}}]{Fabio Bonassi}
received his B.Sc. and M.Sc. degrees cum laude in Automation \& Control Engineering from Politecnico di Milano, Italy, in 2016 and 2018, respectively.
In 2019 his M.Sc. thesis, developed in collaboration with Energy System Research - RSE S.p.A.,  was awarded the “Claudio Maffezoni” prize. From 2019 to 2022 he was a Ph.D. fellow at Politecnico di Milano, during which he joined the European ELO-X program as an Associated Early Stage Researcher, and he received the Ph.D. degree cum laude in Information Technology in early 2023. Since June 2023, he is a Postdoctoral Researcher in machine learning for control at the Uppsala University, Sweden. His main research interests lie in the application of deep learning models for data-driven control. He is also interested in time-series forecasting and Model Predictive Control schemes, with particular application to the power system. His research activities earned him the IFAC Young Author Award, received at SYSID 2021, and the Dimitris N. Chorafas Ph.D. Award in 2023.
\end{IEEEbiography}
\begin{IEEEbiography}[{\includegraphics[width=1in,height=1.25in,clip,keepaspectratio]{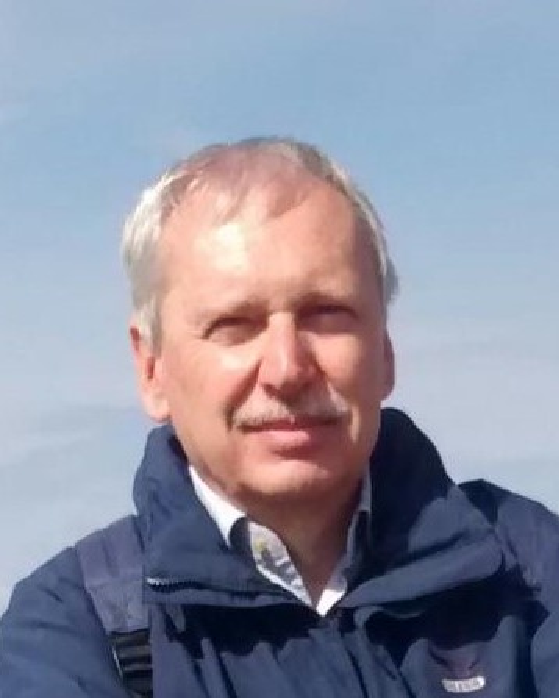}}]{Riccardo Scattolini}
was born in Milano, Italy, in 1956. He is Full Professor of Automatic Control at the Politecnico di Milano. During the academic year 1984/85 he was a visiting researcher at the Oxford University. He has also spent one year working in industry on the simulation and control of chemical plants. He was awarded the national Quazza Premium and the Heaviside Premium of the Institution of  Electrical Engineers, U.K. He has been Associate Editor of the IFAC journal Automatica and of the International Journal of Adaptive Control and Signal Processing. He is author of more than 130 papers published in the main international journals of control, and of more than 150 papers presented at international conferences. His main research interests include modeling, identification, simulation, diagnosis, and control of industrial plants and energy systems, with emphasis on the theory and applications of Model Predictive Control and fault detection methods.
\end{IEEEbiography}
\vfill
\end{document}